\documentclass{aa}  
\usepackage{graphicx}
\usepackage{txfonts}
\usepackage{verbatim,siunitx}
\usepackage[colorlinks=true,linkcolor=blue,urlcolor=blue,citecolor=blue]{hyperref}
\usepackage{xcolor}
\usepackage[normalem]{ulem}
\usepackage{xspace}
\usepackage{booktabs}

\newcommand{\msun}{\ensuremath M_{\odot}}
\newcommand{\rsun}{\ensuremath R_{\odot}}
\newcommand{\lsun}{\ensuremath L_{\odot}}
\newcommand{\eg}{e.g.\@\xspace}
\newcommand{\cf}{cf.\@\xspace}
\newcommand{\ie}{i.e.\@\xspace}
\newcommand{\arepo}{\textsc{arepo}\xspace}
\newcommand{\mesa}{\textsc{mesa}\xspace}

\begin{document}

\title{From 3D hydrodynamic simulations of common-envelope interaction to gravitational-wave mergers}

\titlerunning{CE interactions with a red supergiant primary star} 

\author{Melvin M.~Moreno\inst{1,2}
          \and Fabian R.~N.~Schneider\inst{2,3}\fnmsep\thanks{\email{fabian.schneider@h-its.org}}
          \and Friedrich K.~R{\"o}pke\inst{1,2}
          \and Sebastian T.~Ohlmann\inst{4}
          \and R{\"u}diger~Pakmor\inst{5}
          \and Philipp~Podsiadlowski\inst{6}
          \and Christian~Sand\inst{2}
}

\institute{
    Zentrum f\"ur Astronomie der Universit\"at Heidelberg,
    Institut f\"ur Theoretische Astrophysik, 
    Philosophenweg 12,
    69120 Heidelberg, Germany
\and
    Heidelberger Institut f\"{u}r Theoretische Studien,
    Schloss-Wolfsbrunnenweg 35, 
    69118 Heidelberg, Germany
\and
    Zentrum f{\"u}r Astronomie der Universit{\"a}t Heidelberg,
    Astronomisches Rechen-Institut,
    M{\"o}nchhofstr.\ 12-14, 
    69120 Heidelberg, Germany
\and
    Max Planck Computing and Data Facility, 
    Gie{\ss}enbachstra{\ss}e 2, 
    85748 Garching, Germany
\and
    Max Planck Institute for Astrophysics, 
    Karl-Schwarzschild-Str.\ 1, 
    85748 Garching, Germany
\and
    University of Oxford, St Edmund Hall, Oxford, OX1 4AR, United Kingdom
}

\date{Received YYY; accepted ZZZ}

\abstract{Modeling the evolution of progenitors of gravitational-wave merger
events in binary stars faces two major uncertainties: the common-envelope phase
and supernova kicks. These two processes are critical for the final orbital
configuration of double compact-object systems with neutron stars and black
holes. Predictive one-dimensional models of common-envelope interaction are
lacking and multidimensional simulations are challenged by the vast range of
relevant spatial and temporal scales. Here, we present three-dimensional,
magnetohydrodynamic simulations of the common-envelope interaction of an
initially $10\,\msun$ red supergiant primary star with a black-hole and a
neutron-star companion. Employing the moving-mesh code \arepo and replacing the
core of the primary star and the companion by point masses, we show that the
high-mass regime is accessible to full ab-initio simulations. About half of
the common envelope is dynamically ejected at the end of our simulations and the
ejecta mass fraction keeps growing. Almost complete envelope ejection seems
possible if all ionized gas left at the end of our simulation recombines
eventually and the released energy continues to help unbind the envelope.
We find that the dynamical plunge-in of both companions terminates at orbital
separations too wide for gravitational waves to merge the systems in a Hubble
time. However, the orbital separations at the end of our simulations are
still decreasing such that the true final value at the end of the
common-envelope phase remains uncertain. We discuss the further evolution of
the system based on analytical estimates. A subsequent mass-transfer episode
from the remaining $3\,\msun$ core of the supergiant to the compact companion
does not shrink the orbit sufficiently either. A neutron-star--neutron-star and
neutron-star--black-hole merger is still expected for a fraction of the systems
if the supernova kick aligns favorably with the orbital motion. For double
neutron star (neutron-star--black-hole) systems we estimate mergers in about 9\%
(1\%) of cases while about 77\% (94\%) of binaries are disrupted, \ie supernova
kicks actually enable gravitational-wave mergers in the binary systems
studied here. Assuming orbits smaller by one third after the common-envelope
phase, enhances the merger rates by about a factor of two. Because of the large
post-common-envelope orbital separations found in our simulations, however, a
reduction in predicted gravitational-wave merger events appears possible.}


\maketitle
%

\section{Introduction}
Common-envelope (CE) interaction is one of the major uncertainties in modeling
the progenitor evolution of gravitational-wave induced mergers of neutron stars
(NSs) and black holes (BHs) within the framework of isolated binaries
\citep[\eg][]{voss2003a, dominik2012a, ivanova2013a, tauris2017a,
vigna-gomez2018a, giacobbo2018a, kruckow2018a, belczynski2020a}. In these CE
phases, the compact companion in a binary system is engulfed by the envelope of
a supergiant primary star. Gravitational drag leads to its inspiral. The ``core
binary'' formed from the compact object and the core of the supergiant transfers
orbital energy and angular momentum to the envelope material, which may cause
envelope ejection. This process leaves behind a system with short orbital period
that may eventually merge due to gravitational wave emission, provided it is not
disrupted by the supernova (SN) explosion of the supergiant's core -- which is
the other major uncertainty in the evolution \citep[\eg][]{podsiadlowski2004a,
vigna-gomez2018a, giacobbo2018a, belczynski2020a, mandel2021a}.

Rate predictions for gravitational wave emission events
\citep[\eg][]{abadie2010a, mandel2021b} require the modeling of the source
evolution. Population synthesis studies commonly used for this purpose employ
simple parametrized treatments of CE evolution, see, \eg, \citet{hurley2002a},
\citet{dominik2012a}, \citet{stevenson2017a}, \citet{vigna-gomez2018a},
\citet{mapelli2018a}, \citet{kruckow2018a} and \citet{belczynski2020a}. A
predictive modeling of the entire CE phase also appears difficult in more
elaborate one-dimensional (1D), stellar evolution models, \eg, because the CE
process itself lacks spatial symmetry, especially during the dynamic plunge-in
of the companion into the envelope of the giant star
\citep[\eg][]{bodenheimer1984a, ivanova2013a, ivanova2015a, ivanova2016a,
clayton2017a, fragos2019a}. While these simplified, 1D CE models are very
useful and deliver much insight, a full and proper treatment requires
three-dimensional hydrodynamic simulation approaches.

Even for systems involving low-mass stars such hydrodynamic simulations pose
substantial challenges: The spatial scale range spanned by the extent of the
giant star's envelope, its tiny core, and the -- perhaps compact -- companion
has to be resolved. This wide range of spatial scales introduces a severe
timescale problem because of the Courant--Friedrichs--Lewy (CFL) stability
condition \citep{courant1928a} when using time-explicit hydrodynamics solvers.
Moreover, setting up CE simulations requires diligence to ensure initial
hydrostatic equilibrium (HSE) in the envelope of the primary star. Usually,
these problems are addressed with replacing the stellar cores by point masses
\citep[see, \eg,][for a consistent procedure]{ohlmann2017a}. 
  
In systems involving massive primary stars, the challenges increase even
further: The larger spatial extent of supergiants amplifies the scale problem.
Supergiant stars have a less pronounced core-envelope structure and it is
unclear which part of the core can safely be absorbed in the simplified point
mass treatment \citep[see, \eg,][]{dewi2000a, tauris2001a, podsiadlowski2003a,
kruckow2016a}. For very massive stars, additional physical effects such as
radiation transport become important for the stellar structure
\citep{ricker2019b}, and convection may also play a role, especially after the
dynamical plunge-in \citep[\eg][]{meyer1979a, sabach2017a, wilson2019a,
wilson2020a}. Given these difficulties in simulating CE interaction with massive
stars, there is only a limited number of published models so far
\citep[\eg][]{terman1995a, ricker2019b, law-smith2020a, lau2021a}. They are
clearly challenged by the wide spatial and temporal scale ranges. Some early
models consider the inspiral of a NS into the envelopes of $16\,\msun$ and
$24\,\msun$ stars and conclude that the binary can likely avoid a merger and
thus survive the CE phase if the massive primary star is an evolved red
supergiant \citep{terman1995a}. \citet{ricker2019a} present a test setup with a
very massive primary star ($82.1 \, \msun$ at the onset of CE evolution) and
follow a system with a massive BH companion for a few orbits.
\citet{law-smith2020a} shortcut the scale problem by removing the outer layers
of their $12 \, \msun$ (at zero-age main sequence) primary star trimming its
envelope from the $\sim$$\num{1000} \,\rsun$ it has at the onset of CE evolution
to $10 \, \rsun$ before starting their hydrodynamic simulation with a companion
representing a NS.   

Here, we present simulations with the moving-mesh, (magneto-)hydrodynamic code
\arepo \citep{springel2010a} that follow the inspiral and envelope-ejection
phase of systems with an initially $10\,\msun$ red supergiant (RSG) primary star
and two companion masses, $1.4$ and $5.0\,\msun$, chosen to represent a NS and a
BH, respectively. The NS and BH companions are not resolved and we do not
consider, \eg, accretion processes, jet launching and neutrino cooling from such
objects. Similar simulations of CE interaction in lower-mass systems with the
\arepo code were published by \citet{ohlmann2016a, ohlmann2016b},
\citet{sand2020a}, \citet{kramer2020a}, and \cite{ondratschek2021a}. We
demonstrate that the high-mass star regime is accessible to current
three-dimensional (3D) hydrodynamic simulations comprising the entire CE system
and covering the full inspiral phase. \citet{lau2021a} conduct simulations
similar to ours, but they employ smoothed-particle-hydrodynamics (SPH)
computations of the CE phase and consider an initially $12\,\msun$ star.

In CE simulations with lower-mass primary stars, the final orbital separations
of the remnant core binary are typically too wide compared to observations of
post-CE binaries \citep[see \eg][]{iaconi2017a}. For our simulations with a
massive primary star, we encounter a similar phenomenon: the inspiral on a
dynamic timescale comes to an end and the orbital separations at the end of
the simulations are too wide for gravitational-wave emission to immediately and
robustly result in NS-NS and NS-BH mergers. Hence, we discuss the further
evolution of the remnant binary after CE ejection through another phase of mass
transfer and the second SN \citep[see also][]{tauris2015a, jiang2021a}.

In Sect.~\ref{sect:methods}, we briefly summarize our CE simulation methods and
describe the preparation of the primary star model and the setup of the binary
systems. The results of our 3D hydrodynamic simulations are presented in
Sect.~\ref{sect:simulations} and the further evolutionary modelling of the
post-CE binary up to the compact-object merger stage is considered in
Sect.~\ref{sect:fate}. We discuss our results in Sect.~\ref{sect:discussion} and
conclude in Sect.~\ref{sect:conclusions}.

\section{Methods}
\label{sect:methods}

We follow the procedures developed by \citet{ohlmann2016a}, \citet{ohlmann2017a}
and \citet{sand2020a} for conducting simulations of CE interaction and refer to
these publications for further details. Here, we summarize only a few key
aspects of our modeling. 

Our approach consists of three steps: First, we evolve a 1D model of a massive
star with \mesa version 12115 \citep{paxton2011a, paxton2013a, paxton2015a,
paxton2018a, paxton2019a} until it reaches the desired supergiant stage
(Sect.~\ref{sect:mesa}). This model is then mapped into the moving-mesh code
\arepo \citep{springel2010a}, that was adapted to CE problems
\citep[][]{ohlmann2016a}. Second, the mapped model is relaxed in \arepo to
ensure that the star is in HSE and remains so over several dynamic timescales
(Sect.~\ref{sect:relaxation}). Third, the actual simulation of the interacting
binary system is carried out by adding a companion to the primary-star model
(Sect.~\ref{sect:arepo}). 

In the moving-mesh, magnetohydrodynamics (MHD) code \arepo, we use the
tabulated OPAL equation of state \citep{rogers1996a,rogers2002a}. This requires
a Riemann solver that is compatible with such a general equation of state. For
solving the Riemann problems at the cell interfaces, we employ the HLLD scheme
\citep{miyoshi2005a} with fallback options to the HLL \citep{harten1983a} and
the Rusanov fluxes \citep{rusanov1961a}, as described in \citet{pakmor2011d}.

\subsection{Common-envelope progenitor}
\label{sect:mesa}

Simulating the evolution of CE phases is challenging in itself and becomes
numerically even more difficult with more massive primary stars relevant for the
formation of NS-NS, NS-BH and BH-BH mergers. In more massive stars, the
gravitational potential is deeper, hence hotter temperatures and faster sound
speeds are reached, reducing the CFL-limited integration timesteps. Also the
spatial scales become larger and even radiation transport is important in the
envelopes of very massive stars \citep[$\gtrsim30\,\msun$; see
\eg][]{sanyal2017a, ricker2019b}.

In order to ease the CE computations, we select a primary star that has an
initial mass as low as possible while still forming a NS at the end of its life.
We pick an initially $10\,\msun$ star of solar metallicity
\citep[\eg][]{sukhbold2016a}. The evolution of the star is followed through core
carbon burning to identify a RSG stage at which the star has developed a clear
core-envelope structure and where the binding energy of the envelope,
\begin{align}
    E_\mathrm{bind} = -\int_{M_\mathrm{core}}^{M_\mathrm{surf}}\,\frac{Gm}{r}\,\mathrm{d}m + \alpha_\mathrm{th} \int_{M_\mathrm{core}}^{M_\mathrm{surf}}\, u \,\mathrm{d}m,
    \label{eq:binding-energy}
\end{align}
is lower than the available orbital energy,
\begin{align}
    \Delta E_\mathrm{orb} = -\frac{G M_\mathrm{core} M_2}{2 a_\mathrm{f}} + \frac{G M_1 M_2}{2 a_\mathrm{i}} \approx -\frac{G M_\mathrm{core} M_2}{2 a_\mathrm{f}}.
    \label{eq:orbital-energy}   
\end{align}
In Eqs.~(\ref{eq:binding-energy}) and~(\ref{eq:orbital-energy}), the integrals
are from some inner core boundary $M_\mathrm{core}$ to the surface
$M_\mathrm{surf}$, $m$ is the mass coordinate and $r$ the corresponding radius,
$u$ is the internal energy (including recombination energies), and the factor
$\alpha_\mathrm{th}$ quantifies the fraction of internal energy considered for
in the overall envelope-binding energy. The mass of the primary star at the
onset of CE interaction is $M_1$ and $M_2$ denotes the mass of the companion.
The initial and final orbital separations are $a_\mathrm{i}$ and $a_\mathrm{f}$.
Having an orbital energy exceeding the binding energy of the envelope makes a
successful envelope ejection possible energetically.

A successful CE phase will remove almost the entire hydrogen-rich envelope of
stars and can thus influence their further evolution and final fate. In
particular, losing the envelope of an initially $10\,\msun$ star early on in the
evolution may make it impossible for the star to explode in a supernova and
produce a neutron star. However, this appears unlikely if the envelope is lost
only after core-helium burning \citep{podsiadlowski2004a, schneider2021a}. 

\begin{figure*}
  \centering
  \includegraphics{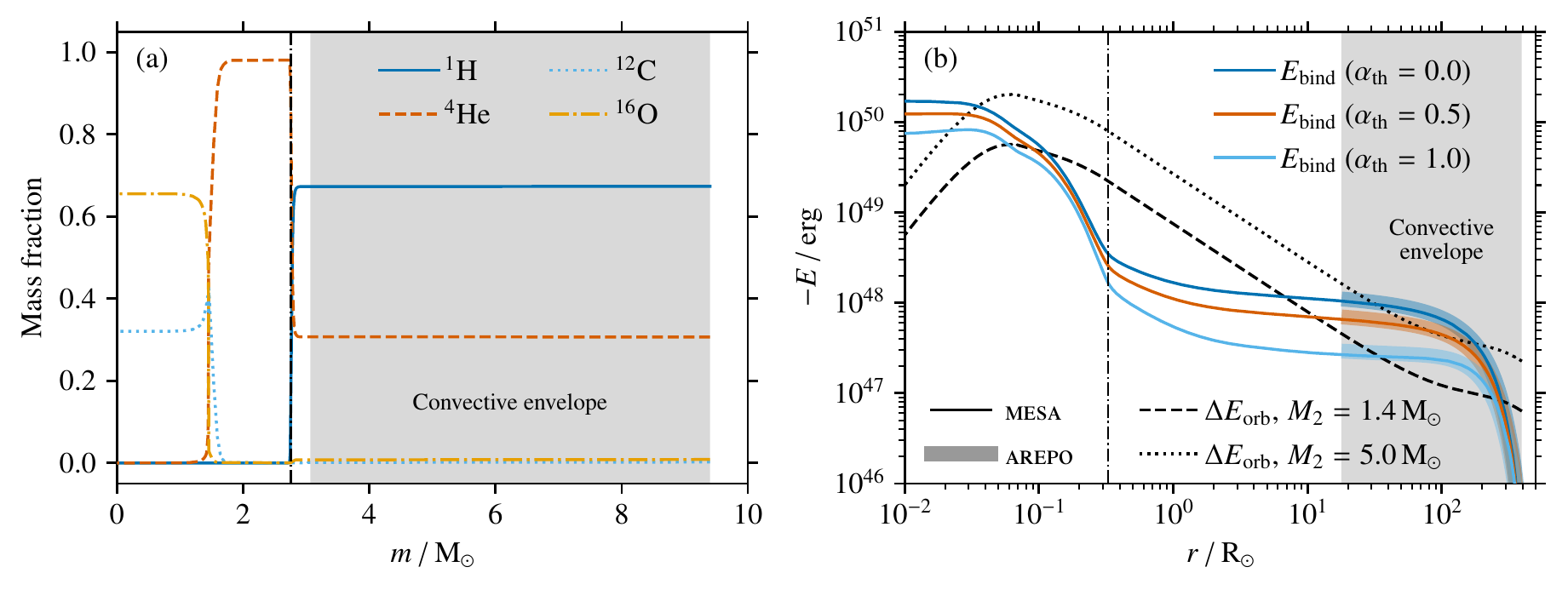}
  \caption{Chemical composition as a function of mass coordinate $m$ (left panel
  a) and envelope binding energy as a function of radius $r$ (right panel b) of
  the RSG primary star. In both panels, the gray shaded area indicates the
  convective envelope and the vertical dot-dashed black line is the location at
  which the hydrogen mass fraction $X=0.1$. In panel (b), also the released
  orbital energy of inspiralling companions of $1.4$ and $5.0\,\msun$ are shown
  (Eq.~\ref{eq:orbital-energy}) as well as the envelope binding energies as
  computed from the relaxed \arepo model (thick transparent lines at $r \geq
  20\,\rsun$; the maximum deviation with respect to the \mesa model is
  ${<}\,5\%$).}
  \label{fig:primary-star}
\end{figure*}

For these reasons, we use a post core-helium-burning star as primary in our CE
simulation that has just reached a radius in excess of its maximum radius before
core-helium burning. At this point in evolution (at an age of about $25.5 \,
\mathrm{Myr}$), the model has a present-day mass of $M_1 = 9.40 \,\msun$, a
radius of $395 \, \rsun$, a logarithmic luminosity of $\log\,L/\lsun = 4.35$,
and an effective temperature of $\num{3564} \, \mathrm{K}$. The chemical
structure of the model and the envelope binding energy are shown in
Fig.~\ref{fig:primary-star}, and the density structure in
Fig.~\ref{fig:structure}.

It is a priori unknown how much of the envelope mass will be ejected in a CE
phase and several criteria are currently being used to determine how much of the
envelope needs to be ejected to avoid an immediate re-growth of the giant star.
The latter is approximated, \eg, by the so-called compression point where the
sonic velocity reaches a maximum \citep[\ie local maximum of the ratio of
pressure and density;][]{ivanova2011a} and by the point in the envelope where
the hydrogen mass fraction $X$ drops below 0.1 \citep[][]{dewi2000a}. Both
points are below the convective envelope (mass coordinate $m=3.07\,\msun$,
radius $r=17.9\,\rsun$) and inside the hydrogen-burning shell
(Fig.~\ref{fig:primary-star}a): the maximum compression point is at
$m=2.78\,\msun$ ($r=0.40\,\rsun$) while the point of $X=0.1$ is at
$m=2.76\,\msun$ ($r=0.33\,\rsun$). So, in our model the two characteristic
points almost coincide.

The envelope binding energy at these points (Fig.~\ref{fig:primary-star}b) is
lower by at least one order of magnitude than the available orbital energy of
binary systems with companion masses of $M_2=1.4\,\msun$ and $5.0\,\msun$. The
binding energies increase only little from the bottom of the convective envelope
to the location at which $X=0.1$ (by factors of 2.3, 2.9 and 5.3 for
$\alpha_\mathrm{th}=0.0$, 0.5 and 1.0, respectively). Contrarily, the available
orbital energy increases by a factor of 48.3. These energetics render a
successful envelope ejection very likely as confirmed by the simulations
discussed in Sect.~\ref{sect:simulations}.

To map the structure of the primary star to the unstructured grid of \arepo, we
follow the procedure developed by \citet{ohlmann2017a}. From the surface of the
star to the core, the density spans more than thirteen orders of magnitude.
Simulating the entire star in \arepo would render the CFL-limited numerical
integration timesteps prohibitively small. Unfortunately, it is not possible at
the moment to cut out only the helium core and have the entire hydrogen-rich
layers (including the $X=0.1$ point) on the \arepo grid because of the wide
range of scales. In our three primary star models, we therefore replace the
innermost $R_\mathrm{cut}=10\,\rsun$, $20\,\rsun$ and $40\,\rsun$ with point
particles of $2.92\,\msun$, $2.97\,\msun$ and $3.03\,\msun$, respectively. These
interact only gravitationally with the envelope gas and the (later added)
companion star. As discussed in Sect.~\ref{sect:discussion-final-orbit}, we find
converged results in terms of orbital separation in our CE simulations for
the models with $R_\mathrm{cut}=10\,\rsun$ and $R_\mathrm{cut}=20\,\rsun$. We
attribute this to the rather slow increase of the binding energy as a function
of radius throughout the hydrogen-rich envelope (Fig.~\ref{fig:primary-star}b)
and the little mass contained in the hydrogen-burning shell ($0.31\,\msun$), \ie
the region above the helium core and below the convective envelope. However,
this is a complex issue and we discuss it in depth in
Sect.~\ref{sect:discussion}.

In the following, $R_\mathrm{cut} = 19.7\,\rsun \approx 20\,\rsun$ is our
default model, \ie the cut is applied very close to the bottom of the convective
envelope, which is located at $17.9\,\rsun$. The innermost $20 \, \rsun$ of the
stellar structure are replaced by a point particle of $M_\mathrm{core} = 2.97 \,
\msun$ and the inner gas cells in the cut-out core region are filled according
to the solution of a modified Lane--Emden equation smoothly blending into the
HSE in the envelope structure \citep{ohlmann2017a}. The point particle together
with the gas cells in the inner region then comprize exactly (deviating only by
1\%) the material of the cut-out core which in the \mesa model has a total mass
of $3.09\,\msun$.

Out of the $2.97 \, \msun$ mass of the core particle, $2.74 \, \msun$ can be
attributed to the carbon-oxygen core and the surrounding helium shell in the
original \mesa model, while the remaining $0.23 \, \msun$ is material that is
chemically part of the hydrogen-rich envelope. We thus include about 97\% of the
hydrogen-rich envelope on the hydrodynamic grid of the simulation.

\subsection{Relaxation}
\label{sect:relaxation}

The mapping from the 1D \mesa model to the coarser hydrodynamic grid in \arepo
introduces discretization errors that perturb the HSE of the envelope and cause
spurious velocities. These are damped out in a relaxation step following
\citet{ohlmann2017a}: Without adding a companion, the primary star is evolved
over ten dynamical timescales (corresponding to about $895 \, \mathrm{d}$).
During the first half of this time, velocities are reduced by a damping term in
the momentum equation that weakens with time, while in the second half the
velocities are no longer artificially suppressed to test the stability of the
resulting structure.

Furthermore, we map the 1D \mesa density structure onto the 3D \arepo grid. This
also introduces a mass discretization error such that the resulting \arepo model
has an approximately 2\% larger total mass of $\approx9.6\,\msun$ compared to
the \mesa target mass of $9.4\,\msun$. When presenting and discussion our CE
simulations, we will thus always refer to the relaxed \arepo model and not the
original \mesa star.

The domain of the \arepo simulation extends far beyond the radius of the primary
star. The cubic simulation box has a length of $3158\,\rsun$ during the
relaxation and $480\,000\,\rsun$ in the CE simulation -- large enough that no
material from the stars is able to reach the boundaries of the simulation box
during our computations. The total number of hydrodynamic grid cells in our
simulation fluctuates around 5 million because of the explicit refinement in
\arepo. Our main refinement criterion is Lagrangian, \ie we aim to keep the mass
of all cells within a factor of two around $1.6\times 10^{-6}\,\msun$. We also
refine a cell if any of its direct neighbours has a volume that is at least a
factor of five smaller than its current volume to avoid abrupt changes in
resolution at steep density gradients. Moreover, we employ explicit refinement
and derefinement to limit the smallest cells to a minimum volume of
$4.2\times10^{29}\,\mathrm{cm}^3$ and the largest cells to a maximum volume of
$3.7\times10^{46}\,\mathrm{cm}^3$ to prevent derefinement of the background
mesh.

Our refinement setting ensures a high spatial resolution close to the core of
the primary star and to the companion while the exterior of the simulated
systems is less well resolved. The resolution is thus highly non-uniform over
the entire simulation domain. Since the vacuum outside the star cannot be
represented in our hydrodynamic simulations, the grid cells outside the stellar
structure are filled with a ``pseudo-vacuum'' consisting of material with a
density of $10^{-15} \, \mathrm{g} \, \mathrm{cm}^{-3}$ and temperature of
$\num{4000} \, \mathrm{K}$.

The gravitational potential of the point particles representing the core of the
primary and the companion is softened out to a radius of $R_\mathrm{soft} =
R_\mathrm{cut}$. To avoid unphysical interaction between the stellar cores when
softened regions overlap, the gravitational softening length is adaptively
reduced as they approach each other such that overlap of the softened
regions is avoided.

\begin{figure}
  \centering
  \includegraphics[width=\linewidth]{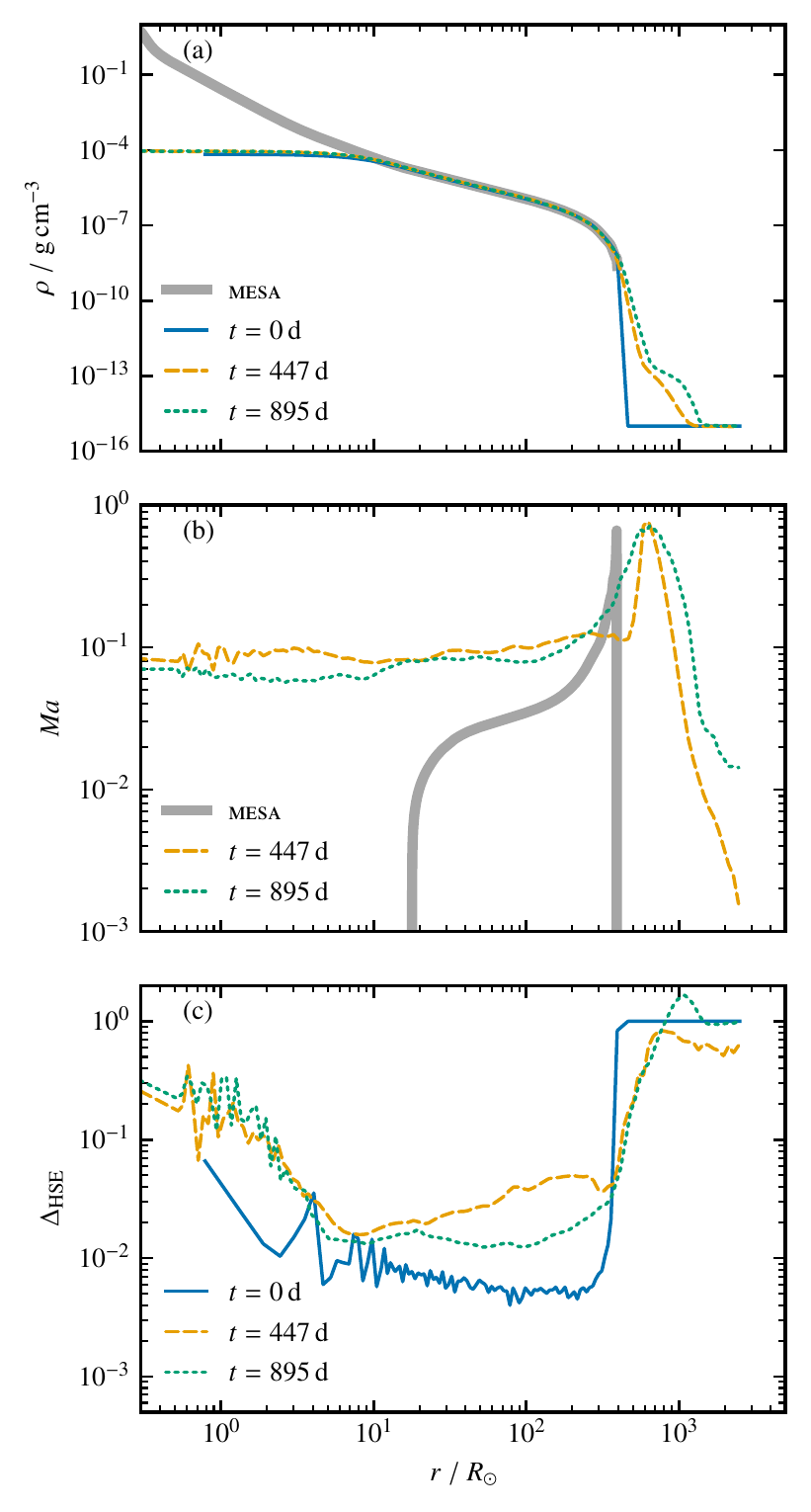}
  \caption{Original \mesa and mapped models of the primary star. Shown are (a)
    density, (b) Mach number and (c) relative deviation from HSE according to
    Eq.~(\ref{eq:hse}) as a function of radius (quantities are averages over
    spherical shells). The times $t = 447 \, \mathrm{d}$ and $t = 895 \,
    \mathrm{d}$ correspond to the end of the damping and the end of the
    relaxation step, respectively. At time $t=0\,\mathrm{d}$, there are no
    velocities in the \arepo model yet, hence the Mach number in panel (b)
    cannot be seen.}
  \label{fig:structure}
\end{figure}

Figure~\ref{fig:structure} illustrates the result of the setup procedure for our
standard stellar model with $R_\mathrm{cut} = 20 \, \rsun$, where we (linearly)
resolve the softening length with $N_\mathrm{cps} = 40$ grid cells exploiting
the adaptive mesh refinement capability of \arepo (``cps'' stands for ``cells
per softening length'', \ie the softened volume is discretized with about $40^3$
cells). The total number of grid cells, $N_\mathrm{cells}$, at the end of the
relaxation is about $4.7$ million. The relaxed model fulfills the stability
criteria set out by \cite{ohlmann2017a}. The Mach numbers $\mathit{Ma}$ after
the relaxation step stay at values around $0.1$ inside the envelope (see
Fig.~\ref{fig:structure}b). This is consistent with the original stellar
structure, where large parts of the resolved envelope are convective (\cf
Fig.~\ref{fig:primary-star}), although the corresponding Mach numbers derived
from the employed mixing length theory are somewhat lower than in the \arepo
model, similar to what \citet{ohlmann2017a} find. Only at the edge of the star
the Mach numbers of both the original and the mapped models approach a value of
one. The lower panel of Fig.~\ref{fig:structure} shows that in the relevant
parts of the envelope there is less than 1\% deviation from HSE, defined as
\begin{align}
    \Delta_\mathrm{HSE} := \frac{|\rho \vec{g} - \nabla P|}{\max (|\rho
    \vec{g}|,|\nabla P|)},
    \label{eq:hse}
\end{align}
where $\rho$, $\vec{g}$, and $P$ denote mass density, gravitational acceleration
and pressure, respectively. The deviations are larger only near the center
($r\lesssim 3\, \rsun$) and in the pseudo-vacuum ($r\gtrsim 400 \, \rsun$),
which is not set up in HSE anyway. This demonstrates that the stellar model is
in an acceptable state of equilibrium before starting the actual binary
interaction simulation. The relaxations of our stellar models with altered
cut radii (but retaining $N_\mathrm{cps} = 40$) give similar results in terms of
stability of the important parts of the envelope.

Since our mapped model cannot completely resolve the steep pressure gradient at
the stellar surface, there is a slight expansion of material (see top panel in
Fig.~\ref{fig:structure}). For consistency with our previous work, we define the
radius of the primary star as that containing 99.9\% of the total mass on the
\arepo grid, which includes some of the pseudo vacuum and yields a radius of
$438 \, \rsun$.

\begin{figure}
  \centering
  \includegraphics[width=\linewidth]{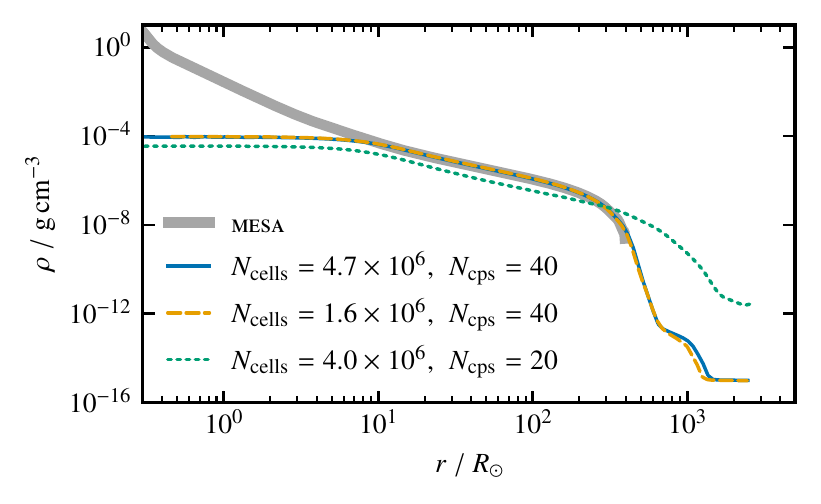}
  \caption{Density profiles of the original stellar model (``\mesa'') and after
    relaxation on \arepo grids with the indicated total number of cells,
    $N_\mathrm{cells}$, and the number of cells per softening length,
    $N_\mathrm{cps}$. Our default model is represented in blue.}
  \label{fig:setup_conv}
\end{figure}

To test the convergence of the initial models with respect to numerical
parameters, we perform two additional simulations. First, we relax our standard
model with $R_\mathrm{cut} = 20 \, \rsun$ to a grid with the total number of
grid cells after the relaxation reduced to $N_\mathrm{cells} = 1.6$ million.
Second, the number of hydrodynamic grid cells resolving the gravitational
softening length of the core particle is reduced to $N_\mathrm{cps} = 20$. The
results are shown in Fig.~\ref{fig:setup_conv}. A reduction of the number of
cells per softening leads to a strong deviation of the density profile after the
relaxation compared with the initial \mesa structure. We conclude that a
sufficiently high resolution of the pressure gradient close to the core particle
is needed to preserve the HSE structure of the original model and perform all
simulations described in the following with $N_\mathrm{cps} = 40$. In contrast,
the total number of grid cells seems to have little impact on the density
structure the \mesa model relaxes into -- at least in the range tested here.
Still, we opt for the higher resolution to improve the representation of the
interaction between the envelope material and the companion and to better
resolve (magneto-)hydrodynamic instabilities.

\subsection{Common-envelope phase}
\label{sect:arepo} 

For the binary interaction simulations, we choose a companion of $M_2 = 5.0 \,
\msun$ representing a BH and a companion of $M_2 = 1.4 \msun$ to represent a NS.
Both are implemented in our model as point masses that interact only
gravitationally, treated in the same way as the core of the primary star. The
companion is placed at an orbital separation of 60\% of that when the RSG
primary star fills its Roche radius. The Roche radius is calculated
approximately according to \citet{eggleton1983a} and the radius of the primary
star after the \arepo relaxation is used ($438 \, \rsun$). We assume corotation
of the giant star's envelope with the binary system and therefore introduce the
corresponding rotational velocity (\ie the envelope is initially in solid-body
rotation). To account for the release of recombination in the expanding envelope
gas, we employ the OPAL equation of state \citep{rogers1996a,rogers2002a} as in
\citet{sand2020a} and \citet{kramer2020a}. We assume all released recombination
energy to thermalize locally and no energy to be lost from dilute parts of the
envelope by radiation. 

We add a dipole magnetic field with polar field strength of
$10^{-6}\,\mathrm{G}$ to the primary star at the beginning of the CE run. The
implementation of the magneto-hydrodynamic solver in \arepo is explained in
\citet{pakmor2011d} and we apply the Powell scheme \citep{powell1999a} for
magnetic-field divergence control \citep[see][]{pakmor2013b}. The seed magnetic
field is strongly amplified in our simulation \citep[\cf][]{ohlmann2016b} and
magnetically-driven, bipolar outflows are observed similar to those in the CE
simulation with a low-mass primary star of \citet{ondratschek2021a}. The
magnetic-field energy saturates at around $10^{45}\,\mathrm{erg}$, and the ratio
of magnetic to kinetic energy never exceeds 3\% in our simulations. The magnetic
fields are irrelevant for the orbital evolution during the dynamic plunge-in of
the CE phase and the ejection of the bulk of the hydrogen-rich envelope, similar
to what is found in lower-mass CE simulations by \citet{ohlmann2016b} and
\citet{ondratschek2021a}. In the future evolution of the remaining binary
systems, the magnetically-driven outflow may affect the binary orbit. However,
there is only little mass left inside the binary orbits at the very end of our
simulations ($\approx10^{-3}\,\msun$), limiting the immediate angular-momentum
loss from the binary systems. These effects will be discussed in a forthcoming
publication.

At the end of our simulations, the deviations of total energy and total angular
momentum compared to the initial states are ${\lesssim}\,0.1\%$ and
${\lesssim}\,1\%$, respectively. A summary of the key initial conditions of
both models at the beginning of the CE runs can be found in
Table~\ref{table:initial-conditions}.

\begin{table}
  \centering
  \caption{Initial conditions of the CE simulations with the NS-like and BH-like companions.}
  \begin{tabular}{l c c}
  \toprule
  CE model & NS-like & BH-like \\
  \midrule
  RSG mass $M_1\,/\,\msun$ & 9.61 & 9.61 \\
  RSG envelope mass $M_\mathrm{env,1}\,/\,\msun$ & 6.65 & 6.65 \\
  RSG radius $R_1\,/\,\rsun$ & 438 & 438 \\
  RSG angular velocity [$\mathrm{s}^{-1}$] & $1.99\times10^{-7}$ & $2.14\times10^{-7}$ \\
  Companion mass $M_2\,/\,\msun$ & 1.4 & 5.0 \\
  Orbital separation $a_\mathrm{ini}\,/\,\rsun$ & 479.0 & 500.8 \\
  Orbital period $P_\mathrm{orb,ini}\,/\,\mathrm{d}$ & 365.8 & 339.5 \\
  Eccentricity $e_\mathrm{ini}$ & 0.0 & 0.0 \\
  Size of simulation box [$\rsun$] & $480{\,}476$ & $501{\,}468$ \\
  Mass of pseudo vacuum [$\msun$] & $1.7\times10^{-3}$ & $2.0\times10^{-3}$ \\
  \bottomrule
  \end{tabular}
  \label{table:initial-conditions}
\end{table}

\section{Results}
\label{sect:simulations}

\subsection{Black-hole mass companion}
\label{sect:sim-bh}

\begin{figure*}
  \centering
  \includegraphics[width=\linewidth]{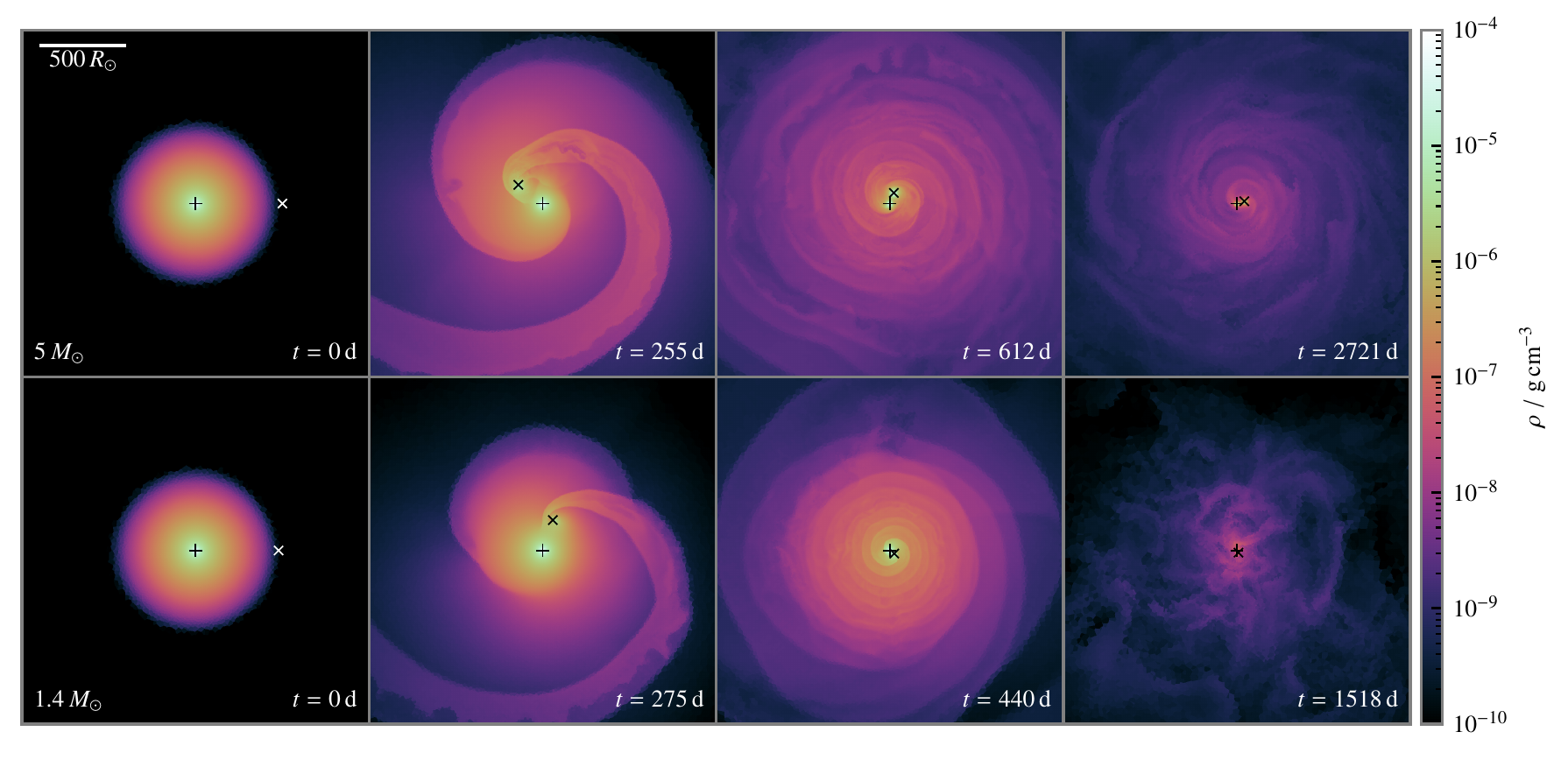}
  \caption{Simulation of CE interaction of a $10 \, \msun$ red giant primary
    star with a BH-mass companion (top row) and with a NS-mass companion (bottom
    row). Shown is the density in slices through the orbital plane at the times
    indicated in the panels. The core of the primary star is marked by a $+$
    symbol; the $\times$ symbol marks the position of the companion. The scale
    bar applies to all snapshots.}
  \label{fig:evol}
\end{figure*}

The evolution of our 3D MHD simulation of CE interaction with a BH-mass
companion (mass ratio $q = M_2/M_1 = 0.53$) is illustrated in the top row of
Fig.~\ref{fig:evol}, and a summary of the final configuration can be found
in Table~\ref{table:final-config}. After setting up the companion at a distance
of $501 \, \rsun$, \ie well above the surface of the primary star, we follow the
system for about six times the initial orbital period -- a duration that
corresponds to $187$ orbits of the progressively tightening core binary system.
Although the stellar envelope and the companion are in corotation initially,
tidal interaction quickly leads to a deformation of the envelope structure,
accumulation of material in the vicinity of the companion, and a drag force
causing the companion to plunge into the envelope. The core of the primary RSG
star and the BH companion orbit each other. Their supersonic motion inside the
envelope leads to the formation of a spiral shock structure easiest visible in
the snapshot taken at $t = 255 \, \mathrm{d}$ in Fig.~\ref{fig:evol}.  By this
time, the companion has completed about one and a half orbits and the separation
between the cores has decreased to less than 50\% of its initial value. Within
the spiral arms, shear flows give rise to instabilities that leave their
imprints on the density structure (see snapshot at $t = 612 \, \mathrm{d}$ in
Fig.~\ref{fig:evol}), similar to what is observed in low-mass systems
\citep{ohlmann2016a}. After $2721 \, \mathrm{d}$, the overall density has
decreased significantly due to envelope expansion (top right panel in
Fig.~\ref{fig:evol}). The spiral structure is still visible, although perturbed
by shear instability. The evolution of the density in the orbital plane is shown
in the video Fig.~\ref{fig:movie-bh}.

\begin{figure}
  \centering
  \includegraphics[width=\linewidth]{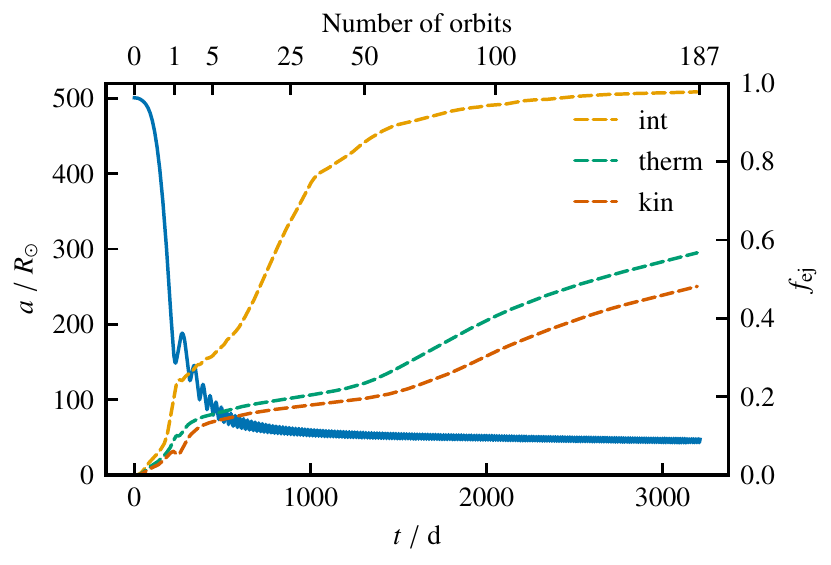}
  \caption{Orbital evolution (solid line, left axis) and mass ejection according
    to the three energy criteria defined in Sect.~\ref{sect:sim-bh} (dashed
    lines, right axis) in the simulation with a BH-mass companion.}
  \label{fig:unbound-bh}
\end{figure}

We now turn to the question of envelope ejection. We determine the mass of
unbound envelope material according to three different energy criteria: gas
inside a hydrodynamic grid cell is considered unbound if its net energy is
positive. In our first, most conservative criterion, we only count gravitational
potential energy and kinetic energy to the net energy of the cell. We will refer
to this as ``kinetic energy criterion''. A  weaker, ``thermal energy criterion''
adds the thermal energy of the gas to the net budget. The third, most optimistic
estimate for envelope unbinding is the ``internal energy criterion'' that counts
in the entire internal energy of the gas including its ionization and radiation
energy. The ionization energy may be released in recombination processes if
envelope expansion moves material below the ionization threshold. We assume
local thermalization of recombination energy and therefore it is bound to
support envelope ejection in our model. Our optimistic internal energy
criterion may overestimate this effect, because it is not guaranteed that
recombination occurs in all parts of the envelope. The possibility that some of
the released recombination energy is radiated (\citealp{grichener2018a}, but see
\citealp{ivanova2018a}) away instead is also not accounted for in our simplified
model.

The evolution of the fraction of ejected envelope mass compared to the initial
envelope mass that is contained on our hydrodynamic grid (\ie 97\% of the
original hydrogen-rich envelope of the \mesa progenitor, see
Sect.~\ref{sect:mesa}),
\begin{align}
    f_\mathrm{ej} (t) = \frac{M_\mathrm{env,\, unbound}(t)}{M_\mathrm{env,\, ini}},
    \label{eq:def-fej}
\end{align}
is plotted in Fig.~\ref{fig:unbound-bh}. After an initial co-evolution of the
fractions measured according to the three criteria in the first orbit,
$f_\mathrm{ej,\, int}$ starts to rise quickly. The unbound mass fractions
determined with the ``kinetic'' and ``thermal'' criteria stagnate around 20\%
from $500 \, \mathrm{d}$ onward. Only after about $1200 \, \mathrm{d}$, the
envelope has expanded sufficiently so that parts of its material fall below the
ionization thresholds of hydrogen and helium. Recombination energy is released
and converted into thermal and ultimately kinetic energy of the envelope gas.
Therefore, $f_\mathrm{ej,\, therm}$ and $f_\mathrm{ej,\, kin}$ start to rise,
but lag behind the evolution of $f_\mathrm{ej,\, int}$. By the time we terminate
the simulation ($3197 \, \mathrm{d}$), we find $f_\mathrm{ej,\, int} = 0.98$,
while $f_\mathrm{ej,\, therm} = 0.57$ and $f_\mathrm{ej,\, kin} = 0.48$ are
still rising steadily. Given this trend, almost complete envelope ejection
appears possible.

We also show the orbital evolution of the system in Fig.~\ref{fig:unbound-bh}.
The separation of the BH-mass companion and the core of the primary star shrinks
by more than $400 \, \rsun$ during the first few orbits in about $500 \,
\mathrm{d}$, after which the orbital decay slows down significantly. The
oscillations apparent in the evolution of the orbital distance are caused by an
eccentricity that develops from the originally circular orbit. The eccentricity
at the end of the simulation is ${\approx}\,0.06$. There is only a small
decrease in the orbital separation over the many orbits between $\num{1000} \,
\mathrm{d}$ and $3197 \, \mathrm{d}$ when we stop the simulation. By this time,
the distance between the core of the primary and the companion has reduced to
$47 \, \rsun$, and it keeps decreasing at a rate of about $1.0\,
\rsun\,\mathrm{yr}^{-1}$. If this rate would stay constant, the orbit would
decay within $47\,\mathrm{yr}$, \ie over the next ${\approx}\,1300$ orbits. This
is well beyond of what can be followed with a code such as \arepo, and signals
the transition from the fast to the slow inspiral process.

The decay rate itself also decreases (Fig.~\ref{fig:unbound-bh}). We thus expect
the orbital separation to settle to a final value once the rest of the envelope
material is expelled completely and the drag forces cease just as is the case in
the CE simulations of lower-mass stars of \citet{ondratschek2021a}. In our
simulations with a massive primary star, however, the CFL-restricted time steps
become prohibitively small to further follow the evolution over many more
orbits.

\begin{table}
  \centering
  \caption{Final configuration of the CE simulations with the NS-like and BH-like companions.}
  \begin{tabular}{l c c}
  \toprule
  CE model & NS-like & BH-like \\
  \midrule
  Time at end of simulation [$\mathrm{d}$] & 1620 & 3197 \\
  Orbital separation $a_\mathrm{f}\,/\,\rsun$ & 15.12 & 47.48 \\
  Orbital period $P_\mathrm{orb,f}\,/\,\mathrm{d}$ & 3.26 & 13.42 \\
  Eccentricity $e_\mathrm{f}$ & 0.011 & 0.059 \\
  Gas mass within $2a_\mathrm{f}$ [$\msun$] & $0.84\times10^{-3}$ & $2.96\times10^{-3}$ \\
  Number of orbits & 309 & 187 \\
  $f_\mathrm{ej,kin}$ & 0.54 & 0.48 \\
  $f_\mathrm{ej,therm}$ & 0.68 & 0.57 \\
  $f_\mathrm{ej,int}$ & $>0.99$ & 0.98 \\
  Orbital decay rate [$\rsun\,\mathrm{yr}^{-1}$] & 0.3 & 1.0 \\
  Magnetic energy [$10^{45}\,\mathrm{erg}$] & 0.5 & 3.1 \\
  Magnetic to kinetic energy ratio & 0.005 & 0.030 \\
  \bottomrule
  \end{tabular}
  \label{table:final-config}
\end{table}

\subsection{Neutron-star mass companion}
\label{sect:sim-ns}

The simulation with a NS-mass companion ($q = 0.15$) and our reference RSG
primary model with $R_\mathrm{cut} = 20 \, \rsun$ was started with an initial
orbital separation of $479 \, \rsun$, again placing the companion above the
stellar surface. The evolution was followed for a duration of about four initial
orbital periods. As visible in the lower panel of Fig.~\ref{fig:evol}, the
hydrodynamic evolution resembles the case of the BH-mass companion, but there
are important differences: The NS-mass companion spirals in faster and deeper
than the more massive BH-mass companion discussed in Sect.~\ref{sect:sim-bh},
because its weaker gravitational force causes less perturbation to the envelope
material. It thus takes longer to expand the envelope and to reduce the drag
force such that the rapid orbital decay comes to an end. When the simulation is
terminated at $1620 \, \mathrm{d}$, the system has completed $309$ orbits of the
tightening core binary system, while the $187$ orbits we follow for the BH-mass
companion case take $3197 \, \mathrm{d}$. 

The deeper inspiral causes a more rapid destruction of spiral shock structure in
the envelope by large-scale instabilities (compare video Fig.~\ref{fig:movie-ns}
with that for the BH-mass companion, Fig.~\ref{fig:movie-bh}). As a result of
the faster and less ordered morphological evolution, the structure in the last
snapshot of the time series in Fig.~\ref{fig:evol} is more dilute in the case of
a NS-mass companion (bottom row) than in the BH-mass companion simulation (top
row). It is also more irregular, and lacks the traces of the previous spiral
shocks imprinted on the envelope material which are still visible in the BH-mass
companion case.

\begin{figure}
  \centering
  \includegraphics[width=\linewidth]{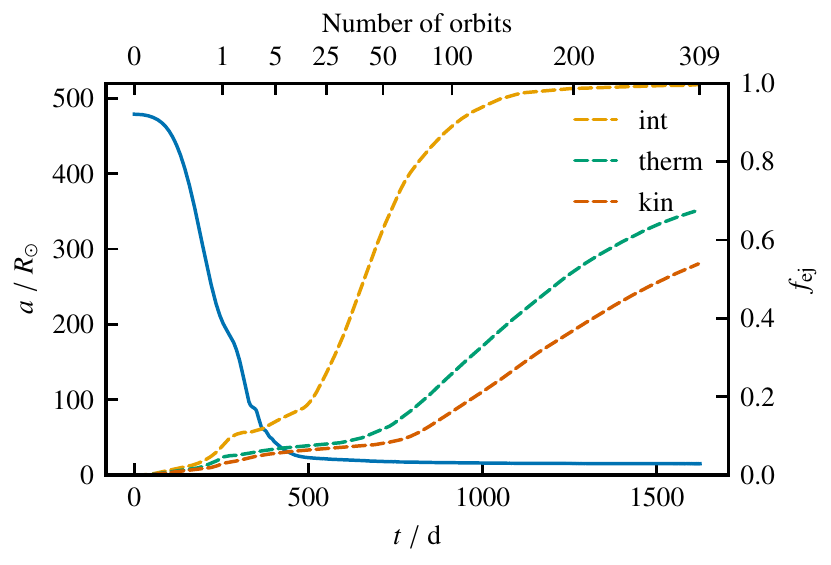}
  \caption{Same as in Fig.~\ref{fig:unbound-bh} but for a NS-mass companion.}
  \label{fig:unbound-ns}
\end{figure}

The evolution of the unbound mass fraction and the orbital decay shown in
Fig.~\ref{fig:unbound-ns} support this interpretation. Compared to the
simulation with the BH-mass companion (Fig.~\ref{fig:unbound-bh}), the inspiral
of the NS-mass object in the first few orbits is deeper and less affected by
eccentricity, which generally seems much less pronounced in the case of the
NS-mass companion. After this initial plunge-in, the core binary system
evolves into a tighter and less eccentric orbit, similar to what is
observed for lower-mass systems (\eg \citealt{sand2020a} find a linear relation
between the mass ratio in the binary system and the ratio of initial to final
orbital separation). At the end of the simulation, the orbital separation is
$15\, \rsun$, and is still decreasing at a rate of about $0.3\,
\rsun\,\mathrm{yr}^{-1}$. The eccentricity at the end is ${\approx}\,0.01$. As
in the BH case (Sect.~\ref{sect:sim-ns}), the decay rate itself is slowing
down, and we are witnessing the transition from the fast to the slow
inspiral process. The entire orbital decay would take $\approx50\mathrm{yr}$
given a constant orbital decay rate, \ie ${\approx}\,5700$ orbits. As shown in
Fig.~\ref{fig:conv} below, however, the decay is not constant but slowly levels
off. Thus, we expect the system to reach a true final orbital separation, but
we cannot follow the evolution for much longer given our CFL-limited time steps.

After about $500 \, \mathrm{d}$, the rate of recombination energy release is
higher in the case of a NS-mass companion than with a BH-mass one. According to
the ``internal energy criterion'' envelope ejection is nearly complete already
by $1000 \, \mathrm{d}$. For the BH-mass companion, ejection takes more than
twice this time. The faster evolution can be attributed to the deeper inspiral
of the NS-mass companion, that releases about 20\% more orbital energy, leading
to a more rapid expansion of the envelope. Shifting larger parts of it quicker
below the ionization threshold, the released recombination energy and its
conversion into thermal and kinetic energy proceeds faster. By the end of our
simulation, $f_\mathrm{ej,\, therm} = 0.68$ and $f_\mathrm{ej,\, kin} = 0.54$
and both are rising steadily. Again, almost complete envelope ejection from the
system seems possible ($f_\mathrm{ej,\,int} > 0.99$).

\subsection{Effective CE ejection efficiencies}
\label{sec:alpha-lambda}

In population-synthesis and other models, the CE phase is often described by an
energy formalism where orbital energy released in the inspiral of the companion
is used to unbind the envelope. Here, the envelope-ejection efficiency,
$\alpha_\mathrm{CE}$, is defined as the ratio of envelope binding energy
($E_\mathrm{bind}<0$; Eq.~\ref{eq:binding-energy}) and the released orbital
energy in the CE phase (Eq.~\ref{eq:orbital-energy}),
\begin{align}
    \alpha_\mathrm{CE}=\frac{E_\mathrm{bind}}{\Delta E_\mathrm{orb}}.
    \label{eq:alpha-ce}
\end{align}
The final orbital separation then is
\begin{align}
    a_\mathrm{f} &= a_i \frac{M_\mathrm{core}}{M_1} \left[ 1 - \frac{1}{\alpha_\mathrm{CE}} \frac{2 E_\mathrm{bind} a_i}{G M_1 M_2} \right]^{-1}, \nonumber \\
    &= a_i \frac{M_\mathrm{core}}{M_1} \left[ 1 + \frac{2}{\alpha_\mathrm{CE} \lambda} \frac{M_\mathrm{env}}{M_2} \frac{a_i}{R_1} \right]^{-1},
    \label{eq:ce-afinal}
\end{align}
where we have introduced the often used, re-defined envelope binding energy in
terms of the so-called $\lambda$ parameter (\cf Eq.~\ref{eq:binding-energy}),
\begin{align}
    E_\mathrm{bind} = -\int_{M_\mathrm{core}}^{M_1}\,\frac{Gm}{r}\,\mathrm{d}m + \alpha_\mathrm{th} \int_{M_\mathrm{core}}^{M_1}\, u \,\mathrm{d}m  \equiv -\frac{G M_1 M_\mathrm{env}}{\lambda R_1}.
    \label{eq:binding-energy-param}
\end{align}
In the latter equation, $M_\mathrm{env}=M_1 - M_\mathrm{core}$ is the envelope
mass and $R_1$ is the radius of the supergiant at the onset of the CE phase.
Both $E_\mathrm{bind}$ and $\lambda$ depend sensitively on how exactly the
envelope binding energy is computed, \ie whether one accounts for thermal and
ionization energy (\ie $\alpha_\mathrm{th}$), and the core-envelope boundary
that sets $M_\mathrm{core}$ (\ie also $M_\mathrm{env}$ and the lower integration
boundary in Eq.~\ref{eq:binding-energy-param}). In this regard,
$E_\mathrm{bind}$ and $\lambda$ are functions of $\alpha_\mathrm{th}$ and
$M_\mathrm{core}$.

The core-envelope boundary setting $M_\mathrm{core}$ is usually interpreted as
the radius down to which the remaining stellar core no longer reacts on
a dynamical timescale with re-expansion to the removal of all above envelope
material such that the CE phase comes to an end. In Sect.~\ref{sect:mesa}, we
show that two often used definitions of this limiting point, the maximum
compression point and the point where $X=0.1$, are virtually identical in our
case.

We now compute the expected final orbital separations applying the energy
formalism in Eq.~\ref{eq:ce-afinal} to the \mesa RSG model as would be done in,
\eg, population synthesis models. Using the maximum-compression point to
separate core and envelope, and assuming $\alpha_\mathrm{th}=1$, we compute the
envelope binding energy of the \mesa RSG model (\cf
Fig.~\ref{fig:primary-star}b). From Eq.~\ref{eq:ce-afinal} and setting
$\alpha_\mathrm{CE}=1$, we then find final orbital separations of
$a_\mathrm{f}\approx 5.9\,\rsun$ and $a_\mathrm{f}\approx 19.2\,\rsun$ for the
NS-mass and BH-mass companion, respectively. These values are \emph{upper}
limits, because we assume a maximum envelope-ejection efficiency of
$\alpha_\mathrm{CE}=1$ and that the entire internal energy of the envelope gas
can be used to unbind the CE ($\alpha_\mathrm{th}=1$). 

These final orbital separations are in contrast to
$a_\mathrm{f}\approx15\,\rsun$ and $a_\mathrm{f}\approx47\,\rsun$ found in our
CE simulations in case of the NS-mass and BH-mass companion, respectively. The
culprit is the a priori unknown core-envelope boundary at which the dynamical
spiral-in stops and that has been estimated at a much deeper location in the RSG
model by the maximum-compression point than what we find in our \arepo
simulation: the drag force on the RSG core and the inspiralling companion is
greatly reduced at much larger separations than predicted by the
maximum-compression point. As shown in Sects.~\ref{sect:sim-bh}
and~\ref{sect:sim-ns}, the orbits still shrink at the end of our simulations,
and it remains uncertain whether improved numerics, different physical setups
and/or yet missing physics would result in smaller final orbital separations in
the \arepo models (see Sect.~\ref{sect:discussion-final-orbit} for a more
detailed discussion on the final orbital separation). In this regard, our final
orbital separations should be interpreted as upper limits.

In order to still obtain a meaningful envelope ejection efficiency for use,
\eg, in population synthesis models, we have to properly adjust the
core-envelope boundary in the energy formalism to the outcome of our CE
simulation. We can then compute the CE ejection efficiency directly from our
\arepo simulation. Using the relaxed \arepo model of the RSG progenitor, we
compute spherical shell averages of internal energy and then integrate the
envelope binding energies following Eqs.~(\ref{eq:binding-energy})
and~(\ref{eq:binding-energy-param}), \cf thick lines in
Fig.~\ref{fig:primary-star}b. From the \arepo progenitor with $R_1=438\,\rsun$,
$M_1=9.61\,\msun$ and $M_\mathrm{core}=2.97\,\msun$ at the onset of the CE
phase, we find $\lambda=0.510$, $0.805$ and $1.911$ for
$\alpha_\mathrm{th}=0.0$, $0.5$ and $1.0$, respectively
(Table~\ref{table:alphas}). 

\begin{table}
    \centering
    \caption{Determined CE ejection efficiencies. The parameters
    $\alpha_\mathrm{th}$, $\lambda$ and $\alpha_\mathrm{CE}$ are defined in
    Eqs.~(\ref{eq:binding-energy}), (\ref{eq:binding-energy-param}) and
    (\ref{eq:alpha-ce}), respectively.}
    \begin{tabular}{c c c c}
    \toprule
     & & NS & BH \\
    $\alpha_\mathrm{th}$ & $\lambda$ & $\alpha_\mathrm{CE}$ & $\alpha_\mathrm{CE}$ \\
    \midrule
    0.0 & 0.510 & 2.29 & 2.60 \\
    0.5 & 0.805 & 1.45 & 1.65 \\
    1.0 & 1.911 & 0.61 & 0.69 \\
    \bottomrule
    \end{tabular}
    \label{table:alphas}
\end{table}

Rearranging Eq.~(\ref{eq:ce-afinal}) for $\alpha_\mathrm{CE}\lambda$, we find
\citep[see \eg][]{webbink1984a, dewi2000a}
\begin{align}
    \alpha_\mathrm{CE}\lambda = \frac{G M_1 M_\mathrm{env}}{R_1} \left[ \frac{G M_\mathrm{core} M_2}{2 a_\mathrm{f}} - \frac{G M_1 M_2}{2 a_\mathrm{i}} \right]^{-1}.
    \label{eq:alpha-lambda}
\end{align}
\emph{Assuming} full envelope ejection (see
Sect.~\ref{sect:discussion-env-ejection}), we get the effective
$\alpha_\mathrm{CE}\lambda$ values from the initial and final orbital
separations in our simulations. In case of the $1.4\,\msun$ NS companion, we
find $\left(\alpha_\mathrm{CE}\lambda\right)_\mathrm{NS}=1.17$ and, from the
simulation with the $5.0\,\msun$ BH companion, we find
$\left(\alpha_\mathrm{CE}\lambda\right)_\mathrm{BH}=1.32$. This translates into
CE ejection efficiencies of $\alpha_\mathrm{CE}=0.61\text{--}2.29$ in the NS
case and $\alpha_\mathrm{CE}=0.69\text{--}2.60$ in the BH case depending on the
choice of $\alpha_\mathrm{th}$ and hence $\lambda$ (Table~\ref{table:alphas}). 

Formally, ejection efficiencies of $\alpha_\mathrm{CE}>1$ indicate that there
are energy sources yet unaccounted for in the energy balance. With the inclusion
of the internal energy of the gas (\ie thermal and ionization energy), we obtain
$\alpha_\mathrm{CE}<1$ for a definition of the core-envelope boundary
representative of our CE simulation (Table~\ref{table:alphas}).

\section{Evolution and final fate of the post-CE binaries}
\label{sect:fate}
  
Both the $1.4\,\msun$ and $5.0\,\msun$ companions can possibly eject most of the
formerly convective, hydrogen-rich envelope of the RSG primary star. The orbital
separations after the dynamical ejection of the CE are still decreasing.
Although we cannot predict the true final orbit accurately in this work
(Sect.~\ref{sect:simulations}), the measured orbital separations are upper
limits. To account for the ongoing orbital decay observed at the end of our
simulations, we also consider scenarios with reduced separations below. The
$2.97\,\msun$ remnant of the RSG star has a remaining lifetime until core
collapse of about $6\times 10^4\,\mathrm{yr}$, which opens several possible
evolutionary paths for the post-CE binary. In the following we assume that the
two companion stars are indeed genuine compact objects, \ie a $1.4\,\msun$ NS
and a $5\,\msun$ BH, and that the entire envelope material of the \arepo models
has been ejected.

Given the mass of the $2.74\,\msun$ helium core of the RSG remnant star and the
$0.23\,\msun$ remaining hydrogen-rich layer, the star is expected to re-expand
again to radii of several tens of $\rsun$ \citep[\eg][]{tauris2017a,
woosley2019a, laplace2020a, vigna-gomez2021a}. This may even happen regardless
of the $0.23\,\msun$ hydrogen-rich layer on top of the helium core
(Sect.~\ref{sect:mesa}). Taking our final orbital separations (periods) of
$15\,\rsun$ ($\approx 3.2\,\mathrm{d}$) and $47\,\rsun$ ($\approx
13.2\,\mathrm{d}$) in case of the NS and BH companion, respectively, another
mass-transfer phase from the RSG remnant onto the compact object appears
inevitable. 

If no mass-transfer phase sets in, the binary system contains a hot helium star
orbiting a compact object. The spectrum of such a hot helium star will be
between an O-type star and a subdwarf, depending on the exact amount of hydrogen
on the stellar surface \citep[][]{gotberg2018a}. Finding such envelope-stripped,
hot helium stars in orbit around other stars is of great interest as they emit
ionising radiation that is thought to be important for, \eg, the reionisation of
the Universe \citep[][]{rosdahl2018a, gotberg2019a, gotberg2020a}.

More likely, however, the RSG remnant will initiate a mass-transfer phase upon
re-expansion. From a binary-evolutionary point-of-view
\citep[\eg][]{dewi2002a, dewi2003a, ivanova2003a, tauris2013a, tauris2015a,
jiang2021a}, mass transfer is most likely stable (for example, see
fig.~3 in \citealt{ivanova2003a} for a $\approx3\,\msun$ donor and an orbital
period of $\approx3.2\,\mathrm{d}$). The aforementioned studies are for
pure helium star donors while our RSG remnant may have a hydrogen-rich surface
layer. Still, given the rather tight post-CE binary orbits, the RSG remnant
cannot expand to such large radii that it can develop a deep convective envelope
and hence the ensuring mass-transfer phase is most likely stable and does
not immediately lead to another CE phase.

Mass transfer will occur in two phases: first, the remaining, hydrogen-rich
layers will be transferred on their thermal timescale. Secondly, a
thermal-timescale mass-transfer episode of helium-rich material sets in. The
timescale of the first mass-transfer phase will be faster. In principle, mass
transfer during this first phase can be so fast that mass accumulates near the
accreting compact object such that it also fills the accretor's Roche volume and
a contact phase could eventually be established. Given the ${\approx}\,3\,\msun$
mass of the RSG remnant, such mass-transferring binaries may be observable as
intermediate-mass X-ray binaries (IMXB). If the mass transfer can bring the
donor mass below ${\approx}\,2\,\msun$ within the remaining lifetime of the
donor star, the system would be classified as a low-mass X-ray binary (LMXB).
However, the duration of mass transfer is of order $10^4\,\mathrm{yr}$, which
makes it difficult to observe such systems.

The accretion onto the NS and BH is likely Eddington-limited,
\begin{align}
    \dot{M}_\mathrm{edd} = \frac{4\pi c R}{\kappa},
    \label{eq:mdot-eddington}
\end{align}
where $R$ is the radius of the compact object, $c$ is the speed of light and
$\kappa$ is the opacity that we approximate by the electron-scattering opacity,
$\kappa_\mathrm{es}=0.2(1+X)\,\mathrm{cm^2}\,\mathrm{g}^{-1}$ and apply a
hydrogen mass fraction of $X=0.7$ by default. Super-Eddington accretion may
also be possible, rendering mass transfer more conservative (see below). For a
NS with radius $10\,\mathrm{km}$ and a non-rotating BH (Schwarzschild radius of
${\approx}\,14.8\,\mathrm{km}$), we find mass accretion rates of
$1.8\times10^{-8}\,\msun\,\mathrm{yr}^{-1}$ and
$2.6\times10^{-8}\,\msun\,\mathrm{yr}^{-1}$, respectively. The rates increase by
70\% for $X=0.0$. Given the remaining lifetime of the RSG remnant, the compact
object can at most accrete some $1\text{--}3\times10^{-3}\,\msun$, \ie a small
amount of mass, but enough to potentially moderately spin it up. 

With Eddington-limited accretion, the mass transfer is highly
non-conservative (typical mass-transfer rates for evolved helium stars are three
to four orders of magnitude larger than the Eddington limit, \eg
\citealt{tauris2015a}). Assuming that a fraction $\beta$ of the mass-transfer
rate $\dot{M}_1$ from the RSG remnant onto the compact object is re-emitted with
the specific orbital angular momentum of the compact object, we can follow the
orbital evolution of the post-CE binary. In this case, the accretion rate of the
compact object is $\dot{M}_2=-(1-\beta)\dot{M}_1$ ($\dot{M}_1<0$). For
$0\leq\beta<1$, we follow \citet{tauris1996a} and, for $\beta=1$, we use
\citep[see also][]{podsiadlowski2002a, postnov2014a}
\begin{align}
    \frac{a}{a_\mathrm{i}} = \left(\frac{M_{1\mathrm{i}}}{M_{1}}\right)^2 \left(\frac{M_{1\mathrm{i}}+M_{2\mathrm{i}}}{M_{1}+M_{2\mathrm{i}}}\right) \exp\left(2\frac{M_{1}-M_{1\mathrm{i}}}{M_{2\mathrm{i}}}\right),
    \label{eq:orbital-evol}
\end{align}
where the index $\mathrm{i}$ denotes values at the onset of the mass-transfer
phase, $a$ is the orbital separation, $M_1$ is the mass of the RSG remnant and
$M_2$ is the mass of the compact object (including the accreted mass). This
solution agrees with that of \citet{tauris1996a} for $\beta\rightarrow1$.
Given these assumptions, $\beta=0$ corresponds to conservative and $\beta=1$
to non-conservative mass transfer.

\begin{figure}
  \centering
  \includegraphics{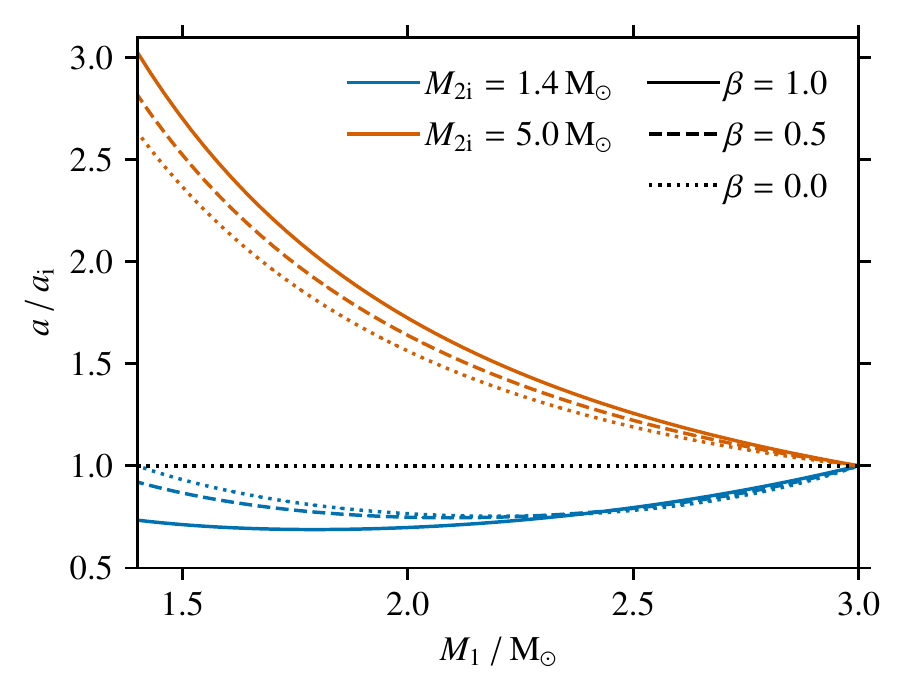}
  \caption{Hypothetical orbital evolution of the remnant binary system. The
  ratio of present-day to initial orbital separation, $a/a_\mathrm{i}$, is
  plotted as a function of the mass of the remnant primary star, $M_1$.
  Evolution is from right to left, \ie from high masses $M_1$ to low masses. The
  initial companion masses are $M_{2\mathrm{i}}=1.4\,\msun$ and $5.0\,\msun$,
  and for each mass three different mass ejection fractions $\beta$ are
  applied.}
  \label{fig:orbital-evolution-mass-transfer}
\end{figure}

The evolution of the orbital separation with respect to its initial value is
shown as a function of donor mass $M_1$ in
Fig.~\ref{fig:orbital-evolution-mass-transfer} for $\beta=1.0$, $0.5$ and
$0.0$. In case of the NS companion of $1.4\,\msun$, the orbit shrinks by at
most 30\% while the orbit always widens in the $5.0\,\msun$ BH case. The orbit
widens by at most a factor of 3 for donor masses larger than $1.4\,\msun$. So
the already relatively large orbital separation of the post-CE system with a
$5\,\msun$ BH only gets larger by mass transfer, making a NS-BH merger induced
by gravitational-wave emission even less likely to occur within a Hubble time.
The orbital shrinkage by 30\% in case of the $1.4\,\msun$ NS companion may
reduce the orbital separation from $15\,\rsun$ to $\approx10\,\rsun$. A NS-NS
merger induced by gravitational-wave radiation may thus be possible depending on
the exact further orbital shrinkage just after the dynamic CE phase (see
Sect.~\ref{sect:sim-ns}) and the supernova explosion of the RSG remnant (see
below).

Assuming a mass-transfer rate of $10^{-4}\,\msun\,\mathrm{yr}^{-1}$
\citep[\eg][]{tauris2015a} and a duration of mass transfer of
$10^4\,\mathrm{yr}$ gives a total ${\approx}\,1\,\msun$ loss from the RSG
remnant. A significant fraction of the helium layer may thus be lost
(Fig.~\ref{fig:primary-star}a) and the star could explode as a so-called
ultra-stripped star with a remaining envelope mass less than $\approx
0.2\,\msun$ \citep[][]{tauris2013a, tauris2015a}. A type Ib, iron core-collapse
SN thus seems most plausible with the possibility of an ultra-stripped SN of low
ejecta mass. Detailed binary evolution models are required to settle this
scenario.

\begin{figure*}
  \sidecaption
  \includegraphics{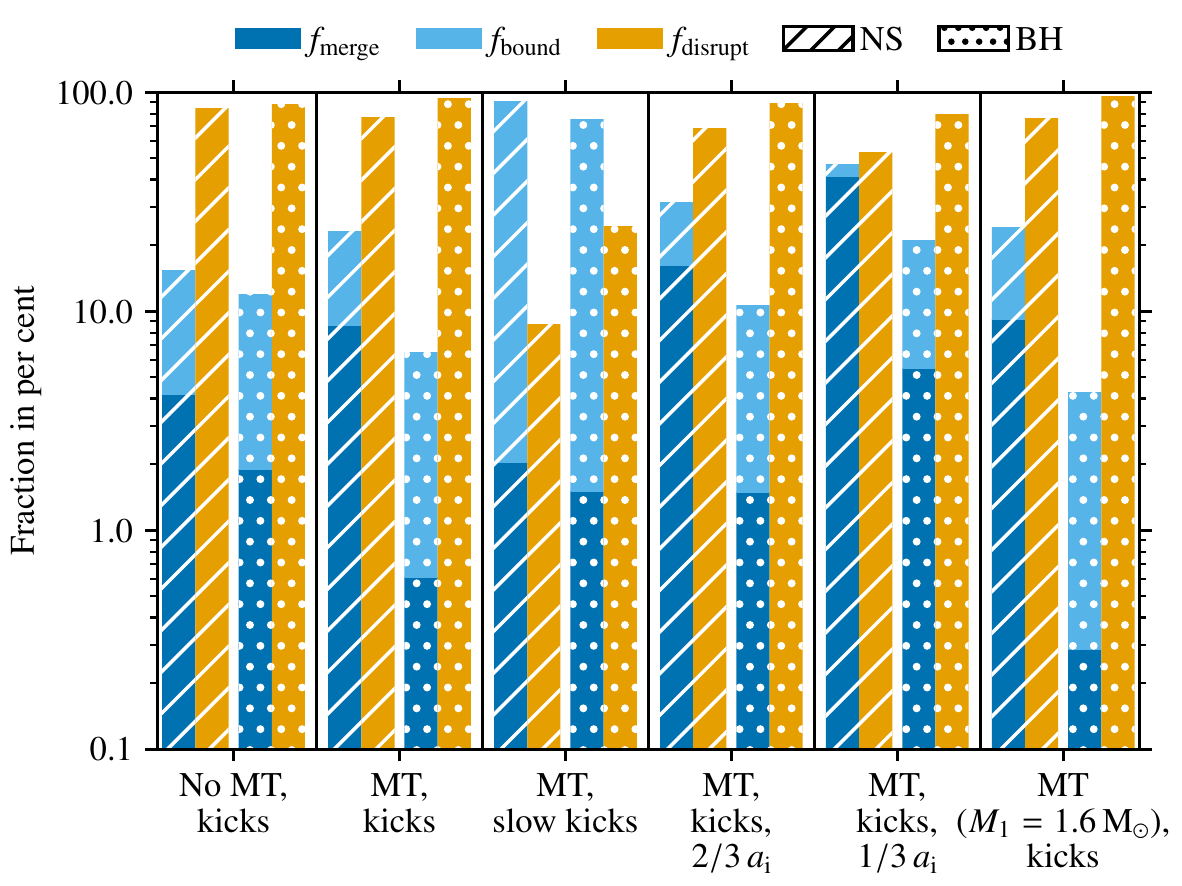}
  \caption{Fraction of systems that merge within a Hubble time,
  $f_\mathrm{merge}$, that remain bound after SN, $f_\mathrm{bound}$, and that
  are disrupted by the SN kick, $f_\mathrm{disrupt}$, in six post-CE
  evolutionary scenarios of binaries with the $1.4\,\msun$ and $5.0\,\msun$
  companions. The six scenarios are as follows: (i) no mass transfer after the
  CE and SN kicks; (ii) our default scenario of non-conservative ($\beta=1.0$)
  mass transfer with final mass $M_1=2.0\,\msun$ and SN kicks; (iii) same as the
  default scenario (ii) but with slower NS kicks from a Maxwellian with
  $\sigma_\mathrm{kick}=50\,\mathrm{km}\,\mathrm{s}^{-1}$; (iv) same as the
  default scenario but assuming an orbital separation after the CE smaller by
  one third; (v) same as (iv) but with an orbital separation smaller by two
  thirds; (vi) same as the default scenario but assuming a final mass of
  $M_1=1.6\,\msun$.}
  \label{fig:sn-kicks-gw-merger}
\end{figure*}

The mass loss in a SN explosion leads to a widening of the orbit and a higher
eccentricity, and the newly-born NS will receive a kick that further perturbs
the orbit. Without a NS kick and assuming that a mass $\Delta M$ is
instantaneously lost from a binary star of masses $M_1$ and $M_2$, the
eccentricity of the orbit is $e=\Delta M / (M_1 + M_2 - \Delta M)$ and the final
orbital separation is
\begin{align}
    \frac{a_\mathrm{f}}{a_\mathrm{i}} = \frac{M_1 + M_2 - \Delta M}{M_1 + M_2 - 2\Delta M}
    \label{eq:final-orbit-after-sn}
\end{align}
if the binary orbit was initially circular. For example, for $M_1=3.0\,\msun$,
$M_2=1.4\,\msun$ and $\Delta M =1.6\,\msun$, \ie assuming that no mass transfer
takes place after the CE event and that there is no NS kick when the
$3.0\,\msun$ star explodes and forms a $1.4\,\msun$ NS, the orbit widens by a
factor of 2.3 and has an eccentricity of about 0.57. In case of the BH companion
of $M_2=5.0\,\msun$, an orbital widening from the SN explosion by a factor of
1.3 and an eccentricity of about 0.25 is expected. Following
\citet{peters1964a}, the GW merger times for of the post-SN binary systems of
$M_1=M_2=1.4\,\msun$, $a=2.3\cdot15\,\rsun\approx34.5\,\rsun$ and $e=0.57$ is
${\approx}\,5500\,\mathrm{Gyr}$, and for $M_1=1.4\,\msun$, $M_2=5.0\,\msun$,
$a=1.3\cdot47\,\rsun\approx61.1\,\rsun$ and $e=0.25$ it is
${\approx}\,38,000\,\mathrm{Gyr}$. So under the assumptions of no mass transfer
and no NS kick, the post-CE binaries will neither lead to a NS-NS nor a NS-BH
merger event within a Hubble time of $13.787\,\mathrm{Gyr}$ \citep{planck2020a}.
Compact-object mergers are possible if the orbital separations at the end of
the CE phase would be smaller than $\approx 3.4\,\rsun$ and $\approx 6.7\,\rsun$
in the NS and BH case, respectively.

Neutron-star kicks may disrupt binaries, but they can also result in tighter
orbital separations and larger eccentricities, \eg if the kick points into the
opposite direction of the motion of the exploding star. In order to quantify how
much NS kicks may help in facilitating a later compact-object merger in our
post-CE binaries, we compute the post-SN orbital separation and eccentricity
following the formalism of \citet{brandt1995a}, specifically using their
equations~(2.4) and~(2.8). A NS of $1.4\,\msun$ is assumed to form from the SN
of the RSG remnant core and we use isotropic NS kicks with kick velocities
following a Maxwellian distribution with
$\sigma_\mathrm{kick}=265\,\mathrm{km}\,\mathrm{s}^{-1}$ \citep{hobbs2005a}.
From the post-SN orbital separation and eccentricity of the non-disrupted binary
systems, we then compute the time it takes the compact objects to merge via
gravitational-wave emission again following \citet{peters1964a}. In
Fig.~\ref{fig:sn-kicks-gw-merger}, we show the fraction of binary systems that
merge within a Hubble time, $f_\mathrm{merge}$, and the fractions of binary
systems that remain bound, $f_\mathrm{bound}$, and that are disrupted by the SN
explosion, $f_\mathrm{disrupt}$.

We consider six scenarios for the evolution after the CE phase to explore the
likelihood of compact-object mergers from the two binary systems studied in
this work. The scenarios are chosen such that the biggest uncertainties in
predicting the most likely final fate of the binary systems are considered.
\begin{enumerate}[(i)]
    \item No mass transfer takes place and the forming NS receives a kick as
    described above. Without NS kicks the binary systems with the NS and BH
    companion do not lead to compact-object mergers within a Hubble time. With
    NS kicks, the binaries disrupt in ${\gtrsim}\,85\%$ of cases, but also lead
    to a NS-NS and NS-BH merger in about 4.2\% and 1.9\% of cases, respectively.
    The merger rates are higher with a NS companion because of the shorter
    orbital separation.
    \item In our default scenario, we assume non-conservative mass transfer with
    $\beta=1.0$ until the donor star reaches a mass of $M_1=2.0\,\msun$ and NS
    kicks as before. Because of the mass transfer, the orbit shrinks in the case
    of the NS companion and widens for the BH companion. This results in a
    larger fraction of bound, post-SN binaries of about 23.1\% in the NS case
    and a smaller fraction of 6.5\% in the BH case when compared to scenario (i)
    without the previous mass-transfer phase. Correspondingly, the NS-NS merger
    fraction increases to 8.7\% and that of NS-BH mergers decreases to 0.6\%.
    \item The mass transfer episode may lead to an ultra-stripped star with a
    small iron core such that the forming NS may receive a slower kick
    \citep[\eg][]{suwa2015a, janka2017a}. For illustrative purposes, we assume a
    Maxwellian kick-velocity distribution with
    $\sigma_\mathrm{kick}=50\,\mathrm{km}\,\mathrm{s}^{-1}$
    \citep[\cf][]{pfahl2002a, podsiadlowski2004a}. Less than 10\% and 15\% of
    binaries with the NS and BH companion, respectively, are disrupted. However,
    the larger fraction of bound binaries does not automatically imply higher
    merger rates. In the NS case, the merger fraction even drops compared to our
    default scenario (ii), because the NS kick is no longer strong enough to
    cause the same fraction of mergers. In the BH case, the merger rate
    increases because the orbital separation at the SN stage is wider and the
    orbital velocity thus smaller such that the NS kick is now more favourable
    to enable mergers within in a Hubble time (NS kicks perturb orbits most if
    they exceed the orbital velocity). This slow-kick scenario becomes more
    relevant the less massive is the RSG remnant at its SN stage. 
    \item The orbital separation at the end of our CE simulations is still
    decreasing and we here assume that it decreases by one third. Mass transfer
    and SN kicks are then treated as in our default scenario (ii). Because of
    the tighter binary orbit, the bound and merger fractions increase. The
    merger rate now is 16.2\% and 1.5\% in the NS and BH cases, respectively.
    \item Decreasing the orbit after the CE phase by two thirds further
    increases the chances for NS-NS and NS-BH mergers to 41.2\% and 5.5\%,
    respectively. 
    \item We assume that the SN takes place when the donor star has reached a
    mass of $M_1=1.6\,\msun$. The SN ejecta mass is now decreased to
    $0.2\,\msun$ compared to $0.6\,\msun$ in the other models, which results in
    a slight increase in the NS-NS merger rate (9.1\%) and a slight decrease in
    the NS-BH merger rates (0.3\%) compared to our default scenario (ii). The
    latter decrease is because of the wider orbit at the end of mass transfer
    with $M_1=1.6\,\msun$ compared to the default scenario with $M_1=2.0\,\msun$
    (see Fig.~\ref{fig:orbital-evolution-mass-transfer}).
\end{enumerate}

In summary, without a NS kick we do not expect NS-NS and NS-BH mergers within a
Hubble time from the binary systems studied here unless there is significant
shrinkage of the orbits directly after the dynamical CE phase. Despite
disrupting most binaries, SN kicks are required to shrink the orbits or lead to
large eccentricities such that compact-object mergers become possible. In our
default scenario, the post-CE binaries have a chance of 8.7\% and 0.6\% to
result in a NS-NS and NS-BH merger within a Hubble time, respectively. The
greatest enhancement in merger rates is found for shorter orbital separations
directly after the dynamical CE phase. Weaker NS kicks actually do not change
the merger fractions drastically but rather only the fraction of bound, double
compact-object binaries. In very rare cases, ${<}\,0.002\%$ and ${<}\,0.02\%$,
the NS is kicked into a bound orbit such that it merges immediately with the
other NS and BH, respectively, on periastron passage. In such rare cases, a
gravitational-wave merger event and short-duration gamma-ray burst may possibly
follow immediately on the SN explosion.

\section{Discussion}
\label{sect:discussion}

\subsection{Envelope ejection}
\label{sect:discussion-env-ejection}

In Sect.~\ref{sec:alpha-lambda}, we estimate the CE ejection efficiency from an
energy formalism. Material in grid cells is considered unbound if the sum of its
gravitational potential, kinetic, and -- in the thermal and internal energy
criteria -- thermal or internal energies is positive. For a number of reasons,
this is only an approximation. 

As discussed in Sect.~\ref{sect:mesa}, not all material down to the $X=0.1$ (or
maximum compression) point is contained on our grid. We cut the envelope
structure at a larger radius and absorb some of the gas into the point particle
representing the core. There is still some gas in the inner hydrodynamic cells,
but it does not properly represent envelope material because it is distributed
over the entire core and the gravitational potential is altered according to the
applied softening. The envelope material between $R_\mathrm{cut}$ and the
$X=0.1$ point has a non-negligible binding energy (Fig.~\ref{fig:primary-star}b)
and would not be removed by the orbital energy released in our simulations.

But also resolved material that is unbound according to our kinetic energy
criterion is not guaranteed to be ejected if it is embedded in bound layers.
Therefore, a better criterion would be based on dynamical considerations, \eg
comparing the radial velocity of the mass element to its escape velocity from
the gravitational potential or an analysis of the final velocity profile to
check whether the envelope is in homologous expansion. Such an analysis should
be carried out once the simulations have been evolved to a stage where the
unbound mass fractions according to the kinetic energy criterion saturate. This,
however, is not the case in the simulations presented here and we therefore
discuss possible envelope ejection results.

By the time we terminate our simulations, nearly complete envelope ejection is
reached for both the NS-mass and BH-mass companion cases according to the
internal energy criterion. The unbound mass fraction according to the kinetic
and thermal energy criteria are still increasing (see Figs.~\ref{fig:unbound-bh}
and \ref{fig:unbound-ns}). This indicates that complete mass ejections according
to the most stringent kinetic energy criterion may eventually be
possible, but this is not guaranteed. It is possible that some part of the
released recombination energy and perhaps also some of the thermal energy is
radiated away instead of aiding envelope ejection (\citealp{grichener2018a},
but also see \cite{ivanova2018a} and for further discussion
\citealp{reichardt2020a} and \citealp{sand2020a}). Such effects, however, are
not accounted for in our model. Radiation transport is not included and local
thermalization of recombination energy is assumed. Future simulations should
implement an appropriate treatment of radiation transport processes in the
envelope. 

The possibility remains that some material stays fully ionised and never
releases its ionization energy (\eg material close to the remnant binary). In
the internal-energy criterion, it is implicitly assumed that also this
ionization energy is available for envelope ejection but it is of course not. In
the NS and BH cases, there is about $0.8\times10^{-3}\,\msun$ and
$3.0\times10^{-3}\,\msun$ of material left close to the binary
(Table~\ref{table:final-config}), respectively, which leads to a negligible
overestimation of the unbound envelope mass. The situation is the same in the
simulation of CE interaction in a system with a low-mass primary star that is
carried out to complete envelope ejection according to the kinetic energy
criterion \citep{ondratschek2021a}: There remains almost no material close to
the binary such that the internal-energy criterion hardly overestimates the
respective envelope ejection efficiency. This is of course at best a
plausibility argument and it emphasizes the need to follow our models for longer
periods of time.

Even if radiation losses or unfavorable deposition of recombination energy
prevent complete envelope ejection in our model, other, currently neglected
effects may still enable it. Treating the companions as point particles is a
crude approximation. In their passages through the envelope, accretion onto the
companions may be possible. Accretion luminosity and perhaps nuclear burning and
the formation of jets may help to expel additional envelope material
\citep[\eg][]{shiber2019a}. However, the results of \citet{macleod2015b}
indicate that the angular momentum barrier due to the density stratification
may prevent efficient mass accretion and the formation of an accretion disk
around compact companions (but see also \citealt{murguia2017a} and
\citealt{chamandy2018a} for cases where accretion is likely possible).
Alternatively, instabilities in the (partly ejected) envelope may be triggered
that can further help eject material \citep[\eg][]{clayton2017a}. Envelope gas
that is not ejected will fall back onto the remnant binary system and parts can
be accreted while other parts may remain in a circumbinary disk. Indeed, such
circumbinary disks are observationally found in, \eg, post-AGB systems that may
have evolved from CE phases \citep[\eg][]{deruyter2006a, vanwinckel2009a,
dermine2013a}.

\subsection{Orbital evolution}
\label{sect:discussion-final-orbit}

\subsubsection{Physical considerations}

We observe an orbital evolution similar to those found in other CE simulations
with lower-mass primary stars \citep[\eg][]{passy2012a, ricker2012a,
ohlmann2016a, iaconi2017a, prust2019a, sand2020a, reichardt2020a,
ondratschek2021a}. After a dynamical plunge-in, that rapidly reduces the
separation of the cores over the first few orbits, the orbital decay slows down
rather abruptly (see Figs.~\ref{fig:unbound-bh} and \ref{fig:unbound-ns}). In
our simulations, this turn-over in the orbital decay occurs at separations far
above the thresholds that would immediately render the considered systems
progenitors of mergers observable in gravitational wave detectors. 

\citet{law-smith2020a} 
take a different approach to find the final orbital separation of a CE phase of
a massive $12\,\msun$ RSG primary and a NS-mass companion. They argue that the
envelope of the RSG primary is easily ejected, and thus strip the RSG primary
down to $10 \, \rsun$ before starting the CE simulation by placing a NS
companion at a distance of $8\, \rsun$ from the primary's center. Our simulation
with a NS companion follows the evolution left out by this shortcut and
indicates that the setup of \citet{law-smith2020a} may not be realistic: The
inspiral of the NS companion in our simulation perturbs the loosely bound outer
parts of the envelope so strongly that the drag forces are unable to reduce the
orbital separation to values chosen by \citet{law-smith2020a} as initial
conditions.

As discussed in Sect.~\ref{sect:discussion-env-ejection}, however, our
models do not capture the innermost envelope material down to the $X=0.1$ point
and we cannot determine whether it would be ejected. Depending on how much of
this mass would need to be ejected, the final orbit would shrink further. A
solid definition of the core of our massive primary star and a hydrodynamic
model that resolves the gas down to this core radius are required to settle this
question.

The rather shallow inspiral after the plunge-in and early CE phase in our
simulations indicates that our considered setups do not produce straight-away
the progenitors of NS-NS and NS-BH that merge in less than a Hubble time. The
observation of merging neutron stars in gravitational-wave detectors
\citep{abbott2017a} raises the question of how reliably our models predict the
evolution of the orbital separation in the core binary system. Interestingly, a
similar problem is encountered in lower-mass systems, where the orbital
separations obtained from simulations are commonly larger than those of observed
post-CE binary systems \citep{iaconi2017a}. Although some observational bias may
play a role in this case, the phenomenon could also indicate a systematic
discrepancy either caused by numerical problems, our inability to evolve CE
simulations long enough or by missing important physical effects in the models.
It is also conceivable that the inspiral results in tighter orbits in less
evolved primary stars with a higher envelope binding energy.

The view of the classical and simplified model of CE evolution is that the
inspiral stalls once a sufficient amount of orbital energy has been released to
expel the envelope material so that the drag on the cores ceases. However, a
simple inspiral at constant orbital-decay rate to a separation where
the difference in orbital energy equals the binding energy of the envelope is
not observed in self-consistent hydrodynamic simulations. The inspiralling
companion rather leads to a complex 3D perturbation of the envelope structure.
The release of recombination energy and its conversion into kinetic energy,
which arguably (see Sect.~\ref{sect:discussion-env-ejection}) is a measure of
how much envelope material has actually been removed from the system, occurs
only after the orbital decay has settled into the slow evolution phase where the
core separation changes only slightly. Thus, ejection of the outer envelope
has little effect on the orbital evolution of the central core binary system.
This picture is supported by simulations comparing CE evolution with and without
taking the release of recombination energy into account. While this process is
fundamental for envelope ejection, it has only a minor effect on the orbital
evolution and the final core separation \citep[\eg][]{sand2020a}.

There are several effects that determine the evolution of the orbital separation
instead. The drag force acting on the cores (recently measured in 3D
hydrodynamic CE simulations by \citealp{chamandy2019a} and
\citealp{reichardt2019a}) depends strongly on the Mach number characterizing
their motion through the envelope gas \cite[\eg][]{ostriker1999a}. It reaches a
maximum at a Mach number of just above unity and drops sharply in the subsonic
regime, which is approached when the companion spirals into the central parts of
the primary star where the sound speed is larger. Moreover, although the
envelope is not unbound at an early stage, it is lifted by the energy injection
lowering the gas density near the cores. Additionally, the drag force is reduced
when the envelope gas spins up and approaches corotation by the action of the
inspiralling companion. At the same time, changing the orbital separation
noticeably requires a drag force on the order of the gravitational force acting
on the cores and determining their orbit. With tightening orbit, the
gravitational force between the cores increases thus reducing the effect of the
decreasing drag force even more. These physical effects seem plausible and all
described effects are in principle accounted for in the physical modeling basis
of our simulations. We therefore argue that the large orbital separations we
observed after the inspiral phase are the correct outcome under the
assumptions and approximations made in our model.

\subsubsection{Numerical considerations}

A remaining concern is that the numerical implementation of our model may
determine the late orbital evolution rather than physics. This concern
originates from our approach to represent the core of the primary RSG star and
the companion as point masses that interact with each other and the gas only via
gravitation. For defining the core particles, the original stellar profile was
cut at a radius $R_\mathrm{cut}$ and the gravitational potential of both point
particles is softened within the same radius. This clearly limits the accuracy
of representation of the stellar profile and the gravitational potential close
to the cores. The impact of the choice of the gravitational softening length
on the results of CE simulations for low-mass systems was addressed by
\citet{reichardt2020a}. While in our simulations the orbital separation of
the system with the BH-mass companion remains well above $R_\mathrm{cut}$, the
turnover in the orbital decay occurs suspiciously close to it in the case of the
NS-mass companion and the final orbital separation of about $15 \, \rsun$ is
even below this radius. This is a reason for concern, because for a cut radius
chosen larger than the core of the giant star, envelope material that should
still contribute to the drag force on the companion is missing at close
distances. 

These arguments highlight the importance of choosing the cut radius as small as
possible, but this implies high computational costs. Capturing parts of the core
on the hydrodynamic grid requires fine spatial resolution and reduces the CFL
time step drastically. Therefore $R_\mathrm{cut} = 20 \, \rsun$ was chosen in
our reference model as a compromise. As can be seen from the top panel of
Fig.~\ref{fig:structure}, this choice may still be acceptable because the cut
density structure starts to deviate rather slowly from the original model below
$R_\mathrm{cut} = 20 \, \rsun$; the difference between the profiles is just
about 1\% at $R= 15 \, \rsun$. Moreover, $R_\mathrm{cut} = 20 \, \rsun$ is only
very slightly above the bottom of the convective envelope
(Fig.~\ref{fig:primary-star}). As described in Sect.~\ref{sect:mesa}, the
compression point and the location where the hydrogen mass fraction is $X=0.1$
are not covered by our simulations. However, there is only little mass between
these bifurcation points and the bottom of the convective envelope
($\approx0.3\,\msun$) and the binding energy of this envelope part is negligible
compared to the available orbital energy (Fig.~\ref{fig:primary-star}b). It
therefore appears that the outcome of our CE simulation might not be much
affected by the inability to resolve the hydrogen envelope in our \arepo runs
down to these points.

\begin{figure}
  \sidecaption
  \includegraphics[width=\linewidth]{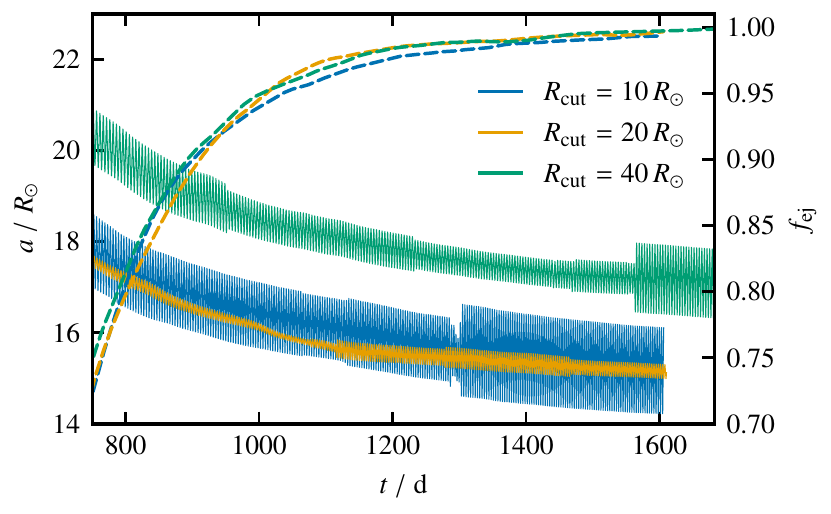}
  \caption{Orbital evolution (solid lines) and mass ejection according to the
    internal energy criterion (dashed lines) for CE simulations with different
    $R_\mathrm{cut}$ in the primary star model.}
  \label{fig:conv}
\end{figure}

In order to asses this more quantitatively, we perform a convergence study
with applying $R_\mathrm{cut} = 10 \, \rsun$ and $R_\mathrm{cut} = 40 \, \rsun$
to the same RSG primary model and our NS-mass companion. The resulting evolution
of the orbital separation between the cores and the unbound mass fraction
according to the internal energy criterion are shown in Fig.~\ref{fig:conv}. For
all three cases, the mass ejection proceeds very similarly pointing to
convergence in this quantity. The orbital evolution and the final separation
between the cores in the model with $R_\mathrm{cut} = 40 \, \rsun$ differs from
the other two models, indicating that convergence is not reached when the core
of the primary star is cut out at this radius in the convective part of the
envelope. However, the models with lower $R_\mathrm{cut}$ agree well in the
average orbital separation. This suggests that the evolution of the distance
between the cores reaches an acceptable level of convergence for
$R_\mathrm{cut} \lesssim 20 \, \rsun$. Since the gravitational softening length
is adaptively reduced when the cores approach each other, the approximate
treatment of gravity inside it is not responsible for the large orbital
separation after rapid plunge-in. The eccentricities, however, show sudden
changes in their temporal evolution and vary considerably. We find final
eccentricities of about 0.06, 0.01 and 0.05 for $R_\mathrm{cut} = 10 \, \rsun$,
$20\,\rsun$ and $40\,\rsun$, respectively. Our models do not provide reliable
results in this quantity. In fact, the eccentricities change when the
gravitational softening lengths around the cores are reduced with tightening
orbits. This dependence on numerical parameters prevents a reliable
determination of the physical eccentricities.

While completing this work, \citet{lau2021a} presented 3D SPH simulations of the
CE phase in systems with a $12\, \msun$ primary star and companions with masses
typical for NSs and BHs. These are largely in agreement with our findings and
support our interpretation.

\section{Conclusions}
\label{sect:conclusions}

We present simulations of CE interaction between a $9.4 \, \msun$ red supergiant
primary star and companions of $1.4\, \msun$ and $5 \, \msun$, typical for NSs
and BHs, respectively. Our results demonstrate that the CE phase in the
massive-star regime is accessible to 3D hydrodynamic simulations. This paves the
way to studying the important yet uncertain CE phases in the progenitor
evolution of compact object mergers observed in gravitational waves. Our
findings can be summarized as follows.

\begin{itemize}
    \item Despite the challenging range of spatial scales, it is possible to
    represent the structure of a $9.4\,\msun$ RSG star on the grid of the
    (magneto-)hydrodynamic code \arepo. This is achieved by replacing the inner
    core (\ie roughly the region below the convective envelope) by a point mass.
    Its associated gravitational softening length is (linearly) resolved with at
    least 40 cells. The resulting stellar model is stable over at least $5$
    dynamic timescales.
    \item We are able to follow the CE evolution with the NS-mass and BH-mass
    companions over 309 and 187 orbits, and find that at least 54\% and 48\%
    of the envelope are unbound, respectively. These fractions are still
    increasing when we terminate our simulations. Almost full envelope ejection
    would be achieved in the optimistic scenario that all remaining
    ionization energy is released by the expanding envelope and can be used for
    unbinding the material.
    \item At the end of our simulations, the final orbital separations
    are $15\,\rsun$ and $47\,\rsun$ in the NS and BH case, respectively,
    and the orbits are still decaying at rates of $0.3$ and
    $1.0\,\rsun\,\mathrm{yr}^{-1}$. We observe a slowing of the orbital decay
    and expect this to come to a complete halt once the entire envelope is
    ejected and there are no drag forces anymore acting on the remnant binary.
    The $1.4\,\msun$ companion spirals in deeper, releases more orbital energy
    and the envelope unbinding proceeds faster than with the more massive
    $5\,\msun$ companion.
    \item Similar to the case of CE evolution with lower-mass primary stars
    \citep{iaconi2017a}, the final orbital separations in our simulations are
    wider than expectedin the sense that without further orbital evolution
    or perturbation, no GW-induced merger is possible for the considered systems
    in a Hubble time. Although simple energy arguments would suggest that
    complete envelope ejection requires a deeper inspiral, the drag force on the
    cores becomes so weak that the orbital decay rate drops by more than two
    orders of magnitude at the end of our simulations. During the plunge-in, the
    orbits shrink at rates of order $1\,\rsun\,\mathrm{d}^{-1}$ while this is
    only $0.3\text{--}1.0\,\rsun\,\mathrm{yr}^{-1}$ in the end.
    \item Within the $\alpha_\mathrm{CE}\lambda$-energy formalism of CE phases
    and assuming full envelope ejection, we find
    $\alpha_\mathrm{CE}\lambda=1.17$ and $\alpha_\mathrm{CE}\lambda=1.32$ in
    case of the NS-mass and the BH-mass companion, respectively. Accounting for
    thermal and ionization energies (\ie $\alpha_\mathrm{th}=1.0$), we find
    $\lambda=1.91$ from the relaxed \arepo model and thus CE envelope-ejection
    efficiencies of $\alpha_\mathrm{CE}=0.61$ and $\alpha_\mathrm{CE}=0.69$ for
    the NS and BH companion, respectively. However, the energy formalism with
    the above quoted parameters only reproduces our simulation outcomes if the
    core-envelope boundary is adjusted to our setup and one does not apply the
    usual $X=0.1$ or maximum compression point.
    \item Although numerical convergence is only indicated but not robustly
    demonstrated, the distances between the stellar cores and the companions
    measured at the end of our simulations provide reasonable estimates within
    our modeling assumptions. Taking these values as upper limits for the
    orbital separations in the post-CE systems, we discuss the likely future
    evolution and final fate as a possible gravitational-wave source. A
    mass-transfer phase from the left-over core of the supergiant onto the
    companion appears likely. The non-conservative mass-transfer episode widens
    (shrinks) the orbit in case of the $5\,\msun$ ($1.4\,\msun$) companion, and
    the ensuing supernova explosion of the RSG core will further affect the
    orbital configuration. The post-CE binary may be visible as an intermediate-
    and low-mass X-ray binary possibly evolving towards an ultra-stripped
    supernova.
    \item Without NS kicks, the resulting NS-NS and NS-BH binaries have final
    orbits too wide to lead to compact-object mergers within a Hubble time
    unless the orbits after the CE phase are smaller by factors of $>$4--7.
    Only with favourably aligned NS kicks that shrink the orbit and/or induce
    high eccentricities, the resulting NS-NS and NS-BH systems lead to
    gravitational-wave merger events in about $8.7\%$ and $0.6\%$ of cases in a
    Hubble time. Tese probabilities are larger by about a factor 2 if the
    orbit right after the CE phase is smaller by one third. So NS kicks
    actually enable gravitational-wave mergers in our case.
\end{itemize}

The most critical parameter determining the merger rate of the post-CE systems
is the orbital separation. Although orbital decay slows down drastically after
the initial plunge-in of the companion, the separation still decreases slightly
when we terminate our simulations. The fact that the slope becomes progressively
shallower suggests that orbital shrinkage will eventually come to an end.
To assess this further and to determine the true final orbital separation,
future simulations should strive to follow the evolution until envelope ejection
is reached according to the kinetic energy criterion \citep[as achieved for
systems with a low-mass primary star by][]{ondratschek2021a}. At the same
time, we are numerically and physically limited, \eg the eccentricities do not
appear to have converged and radiation transport is important to better
understand how much of the ionization energy can indeed be used to help ejecting
the envelope. The accessibility of the massive-star regime for 3D hydrodynamic
CE simulations demonstrated here should be exploited by extending the parameters
space of studied systems to higher masses and different evolutionary stages.

\begin{acknowledgements}
We thank the anonymous reviewer for carefully reading our manuscript and the
useful suggestions that helped improve this work. MMM, FRNS, and FKR acknowledge
support by the Klaus Tschira Foundation. This work has received funding from the
European Research Council (ERC) under the European Union’s Horizon 2020 research
and innovation programme (Grant agreement No.\ 945806). This work is supported
by the Deutsche Forschungsgemeinschaft (DFG, German Research Foundation) under
Germany’s Excellence Strategy EXC 2181/1-390900948 (the Heidelberg STRUCTURES
Excellence Cluster).
\end{acknowledgements}


\begin{appendix} 

\section{Movies of the CE simulations}\label{sec:movies}

\begin{figure}[h!]
	\centering
	\resizebox{\hsize}{!}{\includegraphics{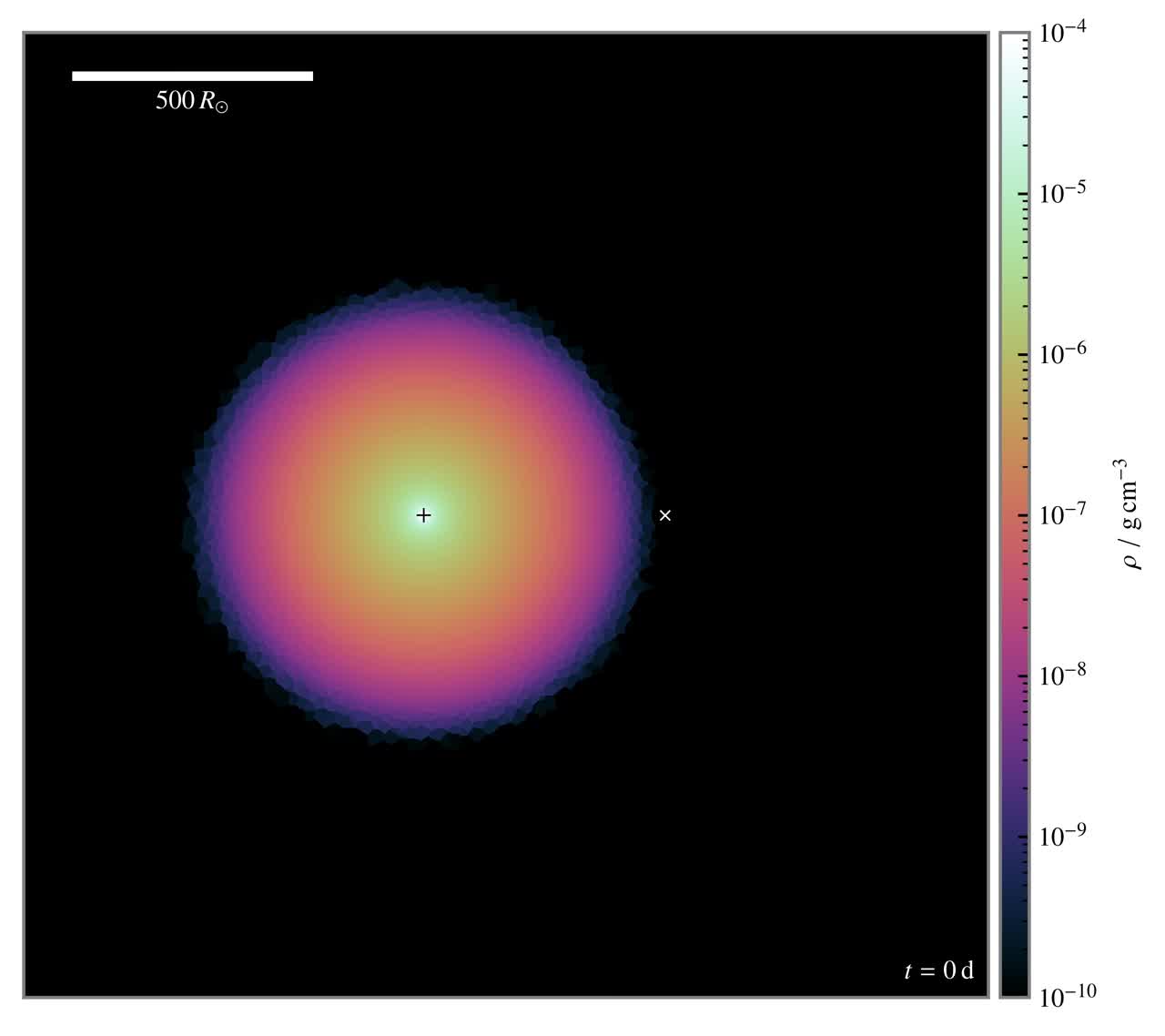}}
	\caption{(\href{https://zenodo.org/record/5721589/files/movie-bh.mp4?download=1}{Movie
	online}) Movie of density evolution in the orbital plane for the simulation
	with a BH-mass companion (marked with $\times$ symbol; the center of the
	primary star is marked with a $+$ symbol). Note that the apparent changes in
	orbital direction result from strobe effects.}
	\label{fig:movie-bh}
\end{figure}

\begin{figure}[h!]
	\centering
	\resizebox{\hsize}{!}{\includegraphics{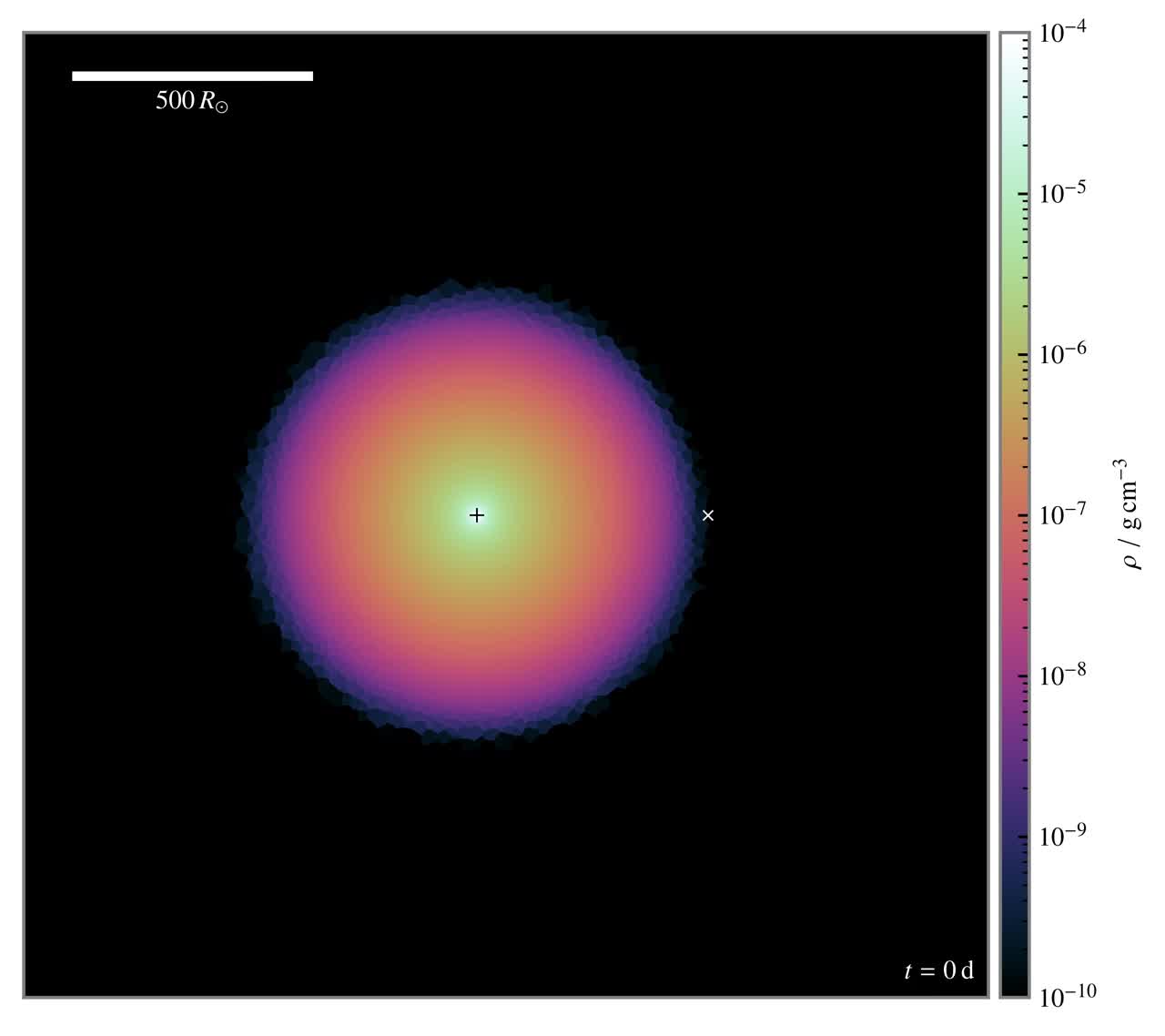}}
	\caption{(\href{https://zenodo.org/record/5721589/files/movie-ns20.mp4?download=1}{Movie
	online}) Same as Fig.~\ref{fig:movie-bh} but for the NS-mass companion.}
	\label{fig:movie-ns}
\end{figure}

Figures~\ref{fig:movie-bh} and~\ref{fig:movie-ns} illustrate the density
evolution in the orbital plane for our two standard simulations with a BH-mass
companion and a NS-mass companion, respectively.

\end{appendix}

\end{document}